\documentclass[reprint, groupedaddress, showpacs, amsmath, amssymb,aps,prd,nofootinbib]{revtex4-1}
\usepackage[bottom]{footmisc} 
\usepackage[utf8]{inputenc}
\usepackage{footnote}
\usepackage{empheq}
\usepackage{amsmath}
\usepackage{mathtools}
\usepackage{amsfonts}
\usepackage{amssymb}
\usepackage{IEEEtrantools}
\usepackage{graphicx}
\usepackage{float}
\usepackage{hyperref}
\usepackage[framemethod=tikz]{mdframed}
\usepackage{lipsum}
\usepackage{tikz-feynman}
\usepackage{slashed}
\usepackage{comment}
\usepackage{xcolor}
\usepackage{natbib}
\usepackage{url}
\usepackage{textcase}
\usepackage{bm}
\usepackage{amsthm}

\tikzfeynmanset{compat=1.1.0}
\newcommand{\p}[4]{\left(#1\right)^{[#2,#3]}_{+(#4)}}
\newmuskip\pFqmuskip
\newcommand\msb{\overline{\text{MS}}}
\newcommand*\pFq[6][8]{%
  \begingroup 
  \pFqmuskip=#1mu\relax
  \mathcode`\,=\string"8000
  \begingroup\lccode`\~=`\,
  \lowercase{\endgroup\let~}\pFqcomma
  {}_{#2}F_{#3}{\left[\genfrac..{0pt}{}{#4}{#5};#6\right]}%
  \endgroup
}
\newcommand{\pFqcomma}{\mskip\pFqmuskip}
\newcommand{\Blue}[1]{\textcolor{black}{#1}}
\newcommand{\Red}[1]{\textcolor{black}{#1}}
\newcommand{\New}[1]{\textcolor{black}{#1}}

\begin{document}
\title{One-Loop Hybrid Renormalization Matching Kernels for Quasi-Parton Distributions}
\author{Chien-Yu Chou}
\email{r09222064@ntu.edu.tw}
\affiliation{Department of Physics and Center for Theoretical Sciences,
National Taiwan University, Taipei, Taiwan 106}
\affiliation{Physics Division, National Center for Theoretical Sciences, Taipei 10617, Taiwan}
\author{Jiunn-Wei Chen}
\email{jwc@phys.ntu.edu.tw}
\affiliation{Department of Physics, Center for Theoretical Sciences,
and Leung Center for Cosmology and Particle Astrophysics,
National Taiwan University, Taipei, Taiwan 106}
\affiliation{Physics Division, National Center for Theoretical Sciences, Taipei 10617, Taiwan}

\begin{abstract}
Large momentum effective theory allows extraction of hadron parton distribution functions  in lattice QCD by 
matching them to quark bilinear matrix elements of hadrons with large momenta.
We calculate the matching kernels for the unpolarized, helicity, and transversity isovector parton distribution functions and skewless generalized parton distributions of all hadrons in the hybrid-RI/MOM scheme. This renormalization scheme uses RI/MOM when the Wilson line length is less than $z_s$, otherwise a mass subtraction scheme is used. By design, the non-hybrid scheme is recovered as $z_s \to \infty$. In the opposite limit, $z_s \to 0$, the self renormalization scheme is obtained. When the parameters $p_z^R=0$ and $\mu^R z_s  \ll 1$, the hybrid-RI/MOM scheme coincides with the hybrid-ratio scheme times the charge of the PDF. We also discuss the subtlety related to the commutativity of Fourier transform and $\epsilon$ expansion in the $\msb$ scheme. 
\end{abstract}
\maketitle
\section{Introduction}

Parton distribution functions (PDFs) describe fundamental structures of hadrons in terms of distributions of quarks and gluons. There are many mid-energy facilities around the world trying to determine these structures and their three-dimensional generalizations such as at Brookhaven and Jefferson Laboratory in the United States,  GSI in Germany, J-PARC in Japan, and a future electron-ion collider (EIC) \cite{AbdulKhalek:2022erw}. The knowledge learned can be applied to searches of physics beyond the Standard Model in energy-frontier experiments like the LHC.

Large-momentum effective theory (LaMET)~\cite{Ji:2013dva,Ji:2014gla} enables computation of the Bjorken-$x$ dependence of hadron PDFs on a Euclidean lattice. It is complementary to experiments especially in kinematic regions that are difficult to access in experiments.
LaMET relates equal-time spatial correlators, whose Fourier transforms are called quasi-PDFs, to PDFs in the limit of infinite hadron momentum.
For large but finite momenta accessible on a realistic lattice, LaMET relates quasi-PDFs to physical ones through a factorization theorem, the proof of which was developed in Refs.~\cite{Ma:2017pxb,Izubuchi:2018srq,Liu:2019urm}.

Since LaMET was proposed, a lot of progress has been made in the theoretical understanding of the formalism~\cite{Xiong:2013bka,Ji:2015jwa,Ji:2015qla,Xiong:2015nua,Ji:2017rah,Monahan:2017hpu,Stewart:2017tvs,Constantinou:2017sej,Green:2017xeu,Izubuchi:2018srq,Xiong:2017jtn,Wang:2017qyg,Wang:2017eel,Xu:2018mpf,Chen:2016utp,Zhang:2017bzy,Ishikawa:2016znu,Chen:2016fxx,Ji:2017oey,Ishikawa:2017faj,Chen:2017mzz,Alexandrou:2017huk,Constantinou:2017sej,Green:2017xeu,Chen:2017mzz,Chen:2017mie,Lin:2017ani,Chen:2017lnm,Li:2016amo,Monahan:2016bvm,Radyushkin:2016hsy,Rossi:2017muf,Carlson:2017gpk,Ji:2017rah,Briceno:2018lfj,Hobbs:2017xtq,Jia:2017uul,Xu:2018eii,Jia:2018qee,Spanoudes:2018zya,Rossi:2018zkn,Liu:2018uuj,Ji:2018waw,Bhattacharya:2018zxi,Radyushkin:2018nbf,Zhang:2018diq,Li:2018tpe,Braun:2018brg,Detmold:2019ghl,Sufian:2020vzb,Shugert:2020tgq,Green:2020xco,Braun:2020ymy,Lin:2020ijm,Bhat:2020ktg,Chen:2020arf,Ji:2020baz,Chen:2020iqi,Chen:2020ody,Alexandrou:2020tqq,Fan:2020nzz,Ji:2020brr,Chen:2018xof,Lin:2018qky,Liu:2018hxv,Liu:2020rqi}.
The method has been applied in lattice calculations of PDFs for the up and down quark content of the  nucleon~\cite{Lin:2014zya,Chen:2016utp,Lin:2017ani,Alexandrou:2015rja,Alexandrou:2016jqi,Alexandrou:2017huk,Chen:2017mzz,Lin:2018pvv,Alexandrou:2018pbm,Chen:2018xof,Alexandrou:2018eet,Lin:2018qky,Liu:2018hxv,Wang:2019tgg,Lin:2019ocg,Liu:2020okp,Lin:2019ocg,Zhang:2019qiq,Alexandrou:2020qtt},
$\pi$~\cite{Chen:2018fwa,Izubuchi:2019lyk,Lin:2020ssv,Gao:2020ito,Gao:2021dbh} and $K$~\cite{Lin:2020ssv} mesons,
and the $\Delta^+$~\cite{Chai:2020nxw} baryon.
Despite limited volumes \cite{Lin:2019ocg,Liu:2020krc} and relatively coarse lattice spacings, previous state-of-the-art nucleon isovector quark PDFs, determined from lattice data at the physical pion mass~\cite{Lin:2018pvv,Alexandrou:2018pbm} and the physical-continuum limit (i.e. with continuum extrapolations at physical pion mass )~\cite{Lin:2020fsj}
have shown reasonable agreement with phenomenological results extracted from the experimental data.
Encouraged by this success, LaMET has also been extended to twist-three PDFs~\cite{Bhattacharya:2020cen,Bhattacharya:2020xlt,Bhattacharya:2020jfj} and generalized parton distributions (GPDs) ~\cite{Bhattacharya:2021oyr},
as well as gluon \cite{Fan:2018dxu,Fan:2020cpa,Salas-Chavira:2021wui},
strange and charm distributions~\cite{Zhang:2020dkn}.
It was also applied to meson distribution amplitudes (DAs)~\cite{Zhang:2017bzy,Chen:2017gck,Zhang:2020gaj,Hua:2020gnw,Hua:2022kcm}
and GPDs~\cite{Chen:2019lcm,Alexandrou:2020zbe,Lin:2020rxa,Alexandrou:2019lfo,Lin:2021brq,Alexandrou:2021bbo}.
Attempts have also been made to generalize LaMET to transverse momentum dependent (TMD) PDFs~\cite{Ji:2014hxa,Ji:2018hvs,Ebert:2018gzl,Ebert:2019okf,Ebert:2019tvc,Ji:2019sxk,Ji:2019ewn,Ebert:2020gxr,Ebert:2022fmh},
to calculate the nonperturbative Collins-Soper evolution kernel~\cite{Ebert:2018gzl,Shanahan:2019zcq,Shanahan:2020zxr,Chu:2022mxh,Shanahan:2021tst}
and soft functions~\cite{Zhang:2020dbb,Li:2021wvl} on the lattice.
LaMET also brought renewed interest in earlier approaches~\cite{Liu:1993cv,Detmold:2005gg,Braun:2007wv,Bali:2017gfr,Bali:2018spj,Detmold:2018kwu,Detmold:2021qln,Liang:2019frk}
and inspired new ones~\cite{Ma:2014jla,Ma:2014jga,Chambers:2017dov,Radyushkin:2017cyf,Orginos:2017kos,Radyushkin:2017lvu,Radyushkin:2018cvn,Zhang:2018ggy,Karpie:2018zaz,Joo:2019jct,Radyushkin:2019owq,Joo:2019bzr,Balitsky:2019krf,Radyushkin:2019mye,Joo:2020spy,Can:2020sxc,HadStruc:2021wmh}.
For recent reviews on these topics, we refer readers to Refs.~\cite{Lin:2017snn,Cichy:2018mum,Zhao:2020vll,Ji:2020ect,Ji:2020byp,Constantinou:2022yye} for more details.

An essential component in the factorization theorem is the matching kernel. The matching kernel compensates 
the difference between the quasi-PDF and PDF in the UV. The matching kernel depends on the renormalization scheme and scale used. While the standard scheme for PDF is $\msb$, several different schemes for quasi-PDF have been proposed. One natural choice is to use the bare quasi-PDF regulated by the lattice 
spacing. However, the kernel has to be recomputed when different versions of lattice discretization are used, and the lattice perturbation theory typically has a slow convergence, not to mention the power divergence in the kernel \cite{Chen:2016fxx}.
Fortunately, using the fact that
the quark bilinear operators are multiplicatively renormalized in coordinate space, different non-perturbative renormalization 
schemes have been developed, such as the regularization-invariant momentum subtraction (RI/MOM) scheme \cite{Stewart:2017tvs,Alexandrou:2017huk} and the ratio scheme \cite{Radyushkin:2017cyf} that will be discussed in detail in this work. A nice feature of these schemes is that now the quasi-PDF is renormalized, 
the dependence on the lattice discretization and slow convergence of lattice perturbation theory can be removed by going to the continuum limit.\footnote{
Ref.~\cite{Zhang:2020rsx} asserted that the continuum limit of RI/MOM for quasi-PDF might not exist. If confirmed, then our hybrid-RI/MOM one-loop kernel in Sec. \ref{h-RM}, which assumes the existence of this continuum limit, will no longer be valid. However, the general procedure on how to convert a non-hybrid matching kernel in momentum space to a hybrid one can still be applied to any other hybrid scheme. } 

Despite the advantage of the RI/MOM and ratio schemes mentioned above, their renormalization factors or counterterms, which, instead of belonging to perturbative UV physics, have non-perturbative IR contributions as well \cite{Ji:2020brr}. To fix this problem, a hybrid scheme was proposed to change the renormalization to a Wilson line mass subtraction scheme when the length of the quark bilinear operator is longer than a scale $z_s \lesssim 0.3$ fm \cite{Ji:2020brr}. 

\Blue{In this manuscript, 
we calculate the matching kernels for the unpolarized, helicity, and transversity isovector PDFs of all hadrons in the hybrid-RI/MOM scheme in Sec. \ref{h-RM}\cite{Ji:2020brr}. (For hadrons with spins less than $1/2$, such as the pions, only the unpolarized PDFs exist.) These matching kernels are identical to those for generalized parton distributions (GPDs) in the zero-skewness limit \cite{Liu:2019urm}. In this limit, GPD has a probability-density interpretation in the longitudinal Bjorken $x$ and the transverse impact-parameter distributions \cite{Burkardt:2002hr} (see also \cite{Ralston:2001xs}). }

\Blue{This hybrid-RI/MOM scheme uses RI/MOM when the Wilson line length $z<z_s$ and a mass subtraction scheme when $z>z_s$. By design, the non-hybrid scheme is recovered as $z_s \to \infty$. In the opposite limit, $z_s \to 0$, the kernel for self renormalization scheme \cite{LatticePartonCollaborationLPC:2021xdx} is obtained and shown in Eq.(\ref{ratio}). Our result suggests that one cannot use self renormalization to all the range of $z$; some modification
of the scheme at small $z$ is needed. A popular limit of hybrid-RI/MOM is to set the parameter $p_z^R=0$ such that the renormalization factor is real. If the parameter $\mu^R$ further satisfies
$\Lambda_{QCD}\ll \mu^R \ll 1/z_s$, then the kernels will coincide with those in the hybrid-ratio scheme (i.e. ratio scheme for $z<z_s$ and 
mass subtraction scheme for $z>z_s$) multiplied by the charges of the PDFs as shown in Sec. \ref{ratio}. }

\Blue{We also discuss the subtlety related to the commutativity of Fourier transform and $\epsilon$ expansion in the $\msb$ scheme in Appendix \ref{A}. This is equivalent to asking whether there is any difference between computing the matching kernel directly using the momentum space Feynman rules and computing in coordinate space then transforming to momentum space. We find that there is an ambiguity in the Fourier transform of the $\ln z^2$ term. If we take the prescription of maintaining quark number conservation to fix the ambiguity, then Fourier transform and $\epsilon$ expansion indeed commute in this case. Ref. \cite{Izubuchi:2018srq} addressed this subtlety not through identifying the ambiguity but by arguing that the terms that caused the non-commutativity would not contribute in the matching provided both the quark and antiquark numbers are finite in the PDF's of hadrons. However, apparently this condition is not always satisfied in global fits. We find that the condition is actually less stringent---only the net quark number needs to be finite, hence is satisfied for all hadrons. }

%

\section{Review of the self and hybrid renormalization schemes} \label{II}

In this section we review the procedure of hybrid renormalization and matching largely following Refs.~\cite{Ji:2020brr,LatticePartonCollaborationLPC:2021xdx}.

We are interested in the quark PDF of a hadron defined through a hadronic matrix element of a quark bilinear operator on a light-cone: \begin{equation}
Q^B(\xi^-,\epsilon)\equiv\frac{1}{2P^+}\langle P\vert\bar{\psi}(\xi^-)\gamma^+W(\xi^-,0)\psi(0)\vert P\rangle,
\end{equation}
where the nucleon momentum is $P^\mu=(P^0,0,0,P^z)$ and the superscripts $\pm$ are the light-cone coordinates $\xi^\pm={(t\pm z)}/{\sqrt{2}}$.
The flavour index is suppressed since we study the non-singlet case where mixing is not considered. The superscripts $B$ and $\epsilon$ indicate that it is a bare matrix element regularized by $d=4-2\epsilon$ dimensional spacetime. 
The Wilson line is a path order (denoted by $P$) line integral of the gauge field $A$
\begin{equation}
W(\xi^-,\rho^-)=P\exp\left(-ig\int^{\xi^-}_{\rho^-}\,d\lambda^- A^+(\lambda)\right),
\end{equation}
which makes sure the quark bilinear is gauge invariant. 
The Fourier transform of the bare light-cone correlator yields a bare quark PDF
\begin{equation}
q^B(x,\epsilon)\equiv\int\frac{d\xi^-P^+}{2\pi}e^{-ix\xi^-P^+}Q^B(\xi^-,\epsilon).
\end{equation}
The $\gamma^+$ structure in the operator indicates the PDF is unpolarized.
We use $q^{\msb}(x,\mu)$ to express the $\msb$ renormalized PDF with the renormalization scale $\mu$.

 The quasi-PDF $\tilde{q}$ is defined as the Fourier transform of an equal time correlator of quark bilinear $\tilde{Q}$ 
\begin{align}
\tilde{q}^B(x,P^z,\epsilon)&\equiv\int\frac{dzP^z}{2\pi}e^{ixP^zz}\tilde{Q}_{\gamma^t}^B(z,P^z,\epsilon)
\end{align}
with equal time correlator along the $z$-direction
\begin{align}
\tilde{Q}_{\gamma^\mu}^B(z,P^z,\epsilon)&=\frac{1}{2P^\mu}\langle P\vert\bar{\psi}(z)\gamma^\mu W(z,0)\psi(0)\vert P\rangle.
\end{align}
Note that the $\mu$ indices on the right hand side are not summed over.
The correlator is multiplicative renormalized \cite{Ishikawa:2017faj,Ji:2017oey,Green:2017xeu} 
\begin{equation}
\tilde{Q}^B(z,P^z,\epsilon)=\tilde{Z}^X(z,P^z,\epsilon,\tilde{\mu})\,\tilde{Q}^X(z,P^z,\tilde{\mu}),\label{7}
\end{equation}
where $\tilde{Z}^X$ is the renormalization factor or counterterm defined in a specific scheme $X$ and $\tilde{\mu}$ is the renormalization scale of the quasi-PDF in the $X$ scheme. This provides a convenient way to convert from one scheme to another.

For a nucleon moving with momentum $P^z$ that is much larger than the nucleon mass $M$ and $\Lambda_{QCD}$, the quasi-PDF can be related to PDF through a factorization theorem. 
In coordinate space, the factorization is \cite{Izubuchi:2018srq}, 
\begin{align}
    &\tilde{Q}^X(z,P^z,\tilde{\mu})\nonumber\\
    =&\int^1_{-1}d\alpha\,\mathcal{C}^X(\alpha,z,\tilde{\mu},\mu)\int^1_{-1}dy\,e^{-i\alpha yzP^z} q^{\msb}(y,\mu)\nonumber\\
    &+\mathcal{O}(z^2 M^2,z^2\Lambda_{QCD}^2) .
    \label{8}
\end{align}
\Blue{To derive this formula, operator product expansion (OPE) is used such that the right hand side of Eq.(\Ref{8}) is a sum of a tower of twist-2 matrix elements with the corresponding Wilson coefficients renormalized in the $X$ scheme plus $\mathcal{O}(z^2\Lambda_{QCD}^2)$ higher twist effects. The twist-2 matrix elements are then further recast into a PDF in coordinate space shown as a Fourier transform of $q^{\msb}$ above.   
} 

In momentum space, the factorization formula is
\begin{align}
\tilde{q}^X(x,P^z,\tilde{\mu})=&\int^1_{-1}\frac{dy}{\vert y\vert}C^X\left(\frac{x}{y},y,\tilde{\mu},\mu,P^z\right)q^{\msb}(y,\mu)\nonumber\\&+\mathcal{O}\left(\frac{M^2}{P_z^2},\frac{\Lambda_{QCD}^2}{P_z^2}\right),\label{9}
\end{align}
where $C^X$ is the matching kernel in the $X$ scheme \cite{Ji:2015qla, Izubuchi:2018srq}.

The relation between the coordinate space and momentum space matching kernels is
\begin{align}
    &C^X\left(\xi,y,\tilde{\mu},\mu,P^z\right)\nonumber\\
    =&\int\frac{P^zdz}{2\pi}e^{i\xi zP^z}\int^1_{-1}d\alpha\,e^{-i\alpha zP^z}\mathcal{C}^X(\alpha,z/y,\tilde{\mu},\mu) .
    \label{10X}
\end{align}
Our main focus in this manuscript is to compute $C^X$ either directly or through $\mathcal{C}^X$. If the scheme $X$ is $\msb$, ratio or RI/MOM, $C^X$ can be computed in both ways. For the hybrid scheme, as we will see, computation through $\mathcal{C}^X$ is more convenient.

The factorization formula in momentum space is rigorously proven only in the $\msb$ scheme \cite{Ma:2014jla,Ji:2020ect}. 
However, we can convert the $\msb$ result to other schemes using the multiplicative renormalization property for the spatial correlators. The procedure is outlined below.  

For an $X$ scheme other than the $\msb$ scheme such as the ratio scheme or the RI/MOM scheme, we can define the conversion factor by the ratio of the correlators
\begin{align}
    Z^X_{\msb}(z,\tilde{\mu},\tilde{\mu'})\equiv\frac{\tilde{Q}^X(z,P^z,\tilde{\mu})}{\tilde{Q}^{\msb}(z,P^z,\tilde{\mu'})}=\frac{\tilde{Z}^{\msb}(z,P^z,\epsilon,\tilde{\mu'})}{\tilde{Z}^{X}(z,P^z,\epsilon,\tilde{\mu})},
    \label{10}
\end{align}
where we have used Eq.(\ref{7}) and the $\epsilon$ dependence in the renormalization factors ought to cancel. The conversion factor also converts the coordinate space matching kernel to a different scheme
\begin{align}
\mathcal{C}^X\left(\alpha,z,\tilde{\mu},\tilde{\mu'},\mu\right)=&Z^X_{\msb}(z,\tilde{\mu'},\tilde{\mu})\mathcal{C}^{\msb}(\alpha,z,\tilde{\mu'},\mu) .\label{12}
\end{align}
Then through Eq.(\ref{10X}), the momentum space matching kernel of scheme $X$ is obtained. 

\subsection{Hybrid scheme}

The conversion factor is a ratio of counterterms which should only have perturbative UV contributions in QCD. However, we will see an example in Sec. \ref{ratio} whose  conversion factor behaves like $1+ c \alpha \ln z^2$, with $c$ a constant. Therefore the conversion factor becomes non-perturbative in the IR or large $z$, which is not desirable.

To fix this problem, a hybrid scheme is proposed \cite{Ji:2020brr}. The idea is to change the renormalization scheme at larger $z$ such that the conversion factor does not grow with $z$. A candidate proposed in \cite{Ji:2020brr} is the Wilson line mass subtraction scheme which argues that the quark bilinear operator has a divergent structure as $C^2 \exp(-\delta m\vert z\vert)$ \cite{Ji:2017oey,Chen:2016fxx,Ishikawa:2017faj,Green:2017xeu}, where $C$ and $\delta m$ are the vertex and mass counterterms that can be determined non-perturbatively using lattice QCD\cite{Green:2020xco,Chen:2016fxx}. Since power divergence is absent in dimensional regularization, the counterterm for the mass subtraction scheme computed in dimensional regularization only gives a $z$ independent constant. So is the counterterm for the $\msb$ scheme. Therefore the conversion factor for the hybrid scheme is a constant for large $z$.

If the conversion factor between the $\msb$ and mass subtraction scheme is simply a constant, perhaps one can use the mass subtraction scheme for the whole range of $z$. As we discussed in Sec. \ref{h-RM} for self renormalization, while the conversion factor is simple,  the matching kernel has a $\ln z^2$ dependence at short distance, hence is not ideal.

Hybrid renormalization is to have the best of both worlds by having mass subtraction at large distance to avoid the large distance logarithm and having RI/MOM or ratio scheme in short distance to avoid the short distance logarithm. The boundary between large and short distance is denoted as $z_s$, which is tested to be $\lesssim 0.3$ fm. Therefore, the conversion factor between the hybrid scheme and the $\msb$ scheme is
\begin{align}
&Z_{\msb}^{\text{hybrid-X}}\left(z,z_s,\tilde{\mu},\tilde{\mu'}\right)\nonumber\\
=&Z_{\msb}^X\left(z,\tilde{\mu},\tilde{\mu'}\right)\theta(z_s-\vert z\vert)+Z_{\msb}^X\left(z_s,\tilde{\mu},\tilde{\mu'}\right)\theta(\vert z\vert-z_s),\label{14}
\end{align}
where the constant conversion factor for $z>z_s$ is fixed by demanding the conversion factor is continuous in $z$.

In the following we called the hybrid scheme using the $X$ scheme in the short $z$ region as ``hybrid-X" scheme, such as the hybrid-ratio and  hybrid-RI/MOM schemes.

\subsection{Self renormalization scheme as a special hybrid scheme}

In Ref. \cite{LatticePartonCollaborationLPC:2021xdx}, an interesting idea was proposed to use the coordinate space correlators of multiple lattice spacings to determine the counterterms. It was called ``self renormalization". 

The self renormalization counterterm is parameterized with explicit dependence on the lattice spacing $a$ \cite{LatticePartonCollaborationLPC:2021xdx}:
\begin{align}
    \tilde{Z}^{\text{self}}(z,a)=&\exp\bigg[\frac{kz}{a\ln(a\Lambda_{QCD})}+m_0z+f(z)a\nonumber\\
    &+\frac{3C_F}{b_0}\ln\left(\frac{\ln(1/(a\Lambda_{QCD}))}{\ln(\mu/\Lambda_{QCD})}\right)\nonumber\\
    &+\ln\left(1+\frac{d}{\ln(a\Lambda_{QCD})}\right)\bigg].
\end{align}
Motivated by the mass subtraction scheme, the first two terms are the linear divergent and finite parts of the mass counterterm, the third term is the discretization error, and the last two terms come from resumming the $z$ independent logarithmic divergence. This form describes the  test data well in Ref. \cite{LatticePartonCollaborationLPC:2021xdx}. At large $z$, no large $\ln z^2$ appears since the logarithmic divergence is $z$ independent. At small $z$, the constant $\msb$ counterterm is also built in. 

The construction of self renormalization is similar to the mass subtraction scheme used in the hybrid renormalization at large $z$. It can be considered as a special case of the hybrid renormalization with $z_s=0$. With the same argument, the conversion factor is a constant
\begin{align}
    &Z_{\msb}^{\text{self}}\left(z,\tilde{\mu},\tilde{\mu'}\right) = g , 
\end{align}
where the constant $g$ can be fixed by the charge of the PDF at $z=0$.

\subsection{Hybrid-ratio scheme as a special case of hybrid-RI/MOM scheme}

In the RI/MOM scheme \cite{Stewart:2017tvs,Alexandrou:2017huk}, the bare coordinate-space matrix element can be renormalized nonperturbatively by demanding that the counterterm cancels all the loop contribution for the matrix element of an off-shell quark state:  
\begin{align}
    &\frac{\tilde{Q}^B_q(z,p^z,\epsilon)}{\tilde{Z}^{\text{RI/MOM}}(z,p^z_R,\epsilon,\mu_R)}\bigg\vert_{p^2=-\mu_R^2,p^z=p^z_R}\nonumber\\
    =&\tilde{Q}^B_q(z,p_R^z,\epsilon \to 0)\bigg\vert_{tree}
    =e^{-ip_R^zz} ,\label{15X}
\end{align}
where a subscript $q$ indicates the matrix element is for a quark state and $p^{\mu}$ is the quark momentum. After renormalization, the UV divergence vanishes so one can talk $\epsilon \to 0$. Here we study the non-singlet quasi-PDF so there is no mixing to the gluon quasi-PDF. 
The UV divergence appears in the Wilson coefficients of an OPE and is independent of the external state, hence one can choose to evaluate the counterterm using a quark state. The quark off-shellness $\mu_R^2 \gg \Lambda^2_{\text{QCD}}$, hence the counterterm $\tilde{Z}^{\text{RI/MOM}}$ can be computed perturbatively. 
The off-shellness introduces a gauge dependence in the matrix element and typically Landau gauge is employed on the lattice. However, as we argue below Eq.(\ref{18}), 
the off-shellness drops out when $\mu_R^2 z^2 \to 0$---a condition employed in the hybrid-RI/MOM scheme. Also, the offshellness behaves like an effective mass, hence the $\gamma^t$ operator is no longer protected by chiral symmetry. A projection to the $\gamma^t$ structure is implied in Eq.(\ref{15}).


In the hybrid-RI/MOM scheme, RI/MOM is applied for $z<z_s$. If $p^z_R=0$, then
\begin{align}
    \frac{\tilde{Q}^B_q(z,0,\epsilon)}{\tilde{Z}^{\text{RI/MOM}}(z,0,\epsilon,\mu_R)}\bigg\vert_{p^2=-\mu_R^2,z<z_s}=1 . 
\end{align}
If we also have $z_s\mu_R\ll1$, then similar to Eq.(\ref{15X}),
\begin{align}
    \tilde{Z}^{\text{RI/MOM}}(z,0,\epsilon,\mu_R)
    =&\tilde{Q}^B_q(z ,0,\epsilon)\bigg\vert_{z \mu_R \to 0}+\mathcal{O}(\mu_R^2 z^2) \nonumber \\
    =&\frac{\tilde{Q}^B(z ,0,\epsilon)}{g}\bigg\vert_{z \mu_R \to 0}+\mathcal{O}(\mu_R^2 z^2) \nonumber \\
    =&\frac{\tilde{Z}^{\text{ratio}}(z,\epsilon)}{g}+\mathcal{O}(\mu_R^2 z^2) .
    \label{18}
\end{align}
\Blue{In the first line of Eq.(\ref{18}), the dimensionless $\tilde{Z}^{\text{RI/MOM}}$ depends on $z$, $\mu_R$ and $\Lambda_{\text{QCD}}$. $\Lambda_{\text{QCD}}$ can be dropped since $\mu_R^2 \gg \Lambda^2_{\text{QCD}}$. Therefore the only dimensionless combination is $z \mu_R$.  \New{The $z \mu_R \to 0$ limit implies
(a) In momentum space, the parton momentum $p^z \gg \mu_R$. The system is effectively on-shell because the $\mu_R$ off-shellness is negligible for highly relativistic partons. 
(b) In coordinate space, the quark matrix element is effectively a local matrix element because $z \ll 1/\mu_R$. }
This matrix element is related to a local
hadron matrix element in the second line with $g$ the charge of the PDF. For the unpolarized PDF, $g=1$. In the last line, we use the definition of the ratio scheme to rewrite the zero momentum hadron matrix element as the counterterm. }

\Blue{The above arguments show that the hybrid-RI/MOM scheme will be, up to a charge, identical to the hybrid-ratio scheme in the limits of
\New{\begin{equation}
p^z_R=0 , \ \ z \le z_s \ll \frac{1}{\mu_R} \ll \frac{1}{\Lambda_{\text{QCD}}} . 
\end{equation}
This connection was first mentioned in passing in Ref. \cite{Ji:2020brr}.  
We have computed the hybrid-ratio kernels for proton isovector PDFs and 
explicitly verified that they can be reproduced by taking $p_R^z = 0$ and $\mu_R \ll p^z$ to their corresponding
hybrid-RI/MOM kernels. The gauge dependence disappears under this limit as expected.}
}
\begin{widetext}

\section{Matching factor of quasi-pdf in the hybrid scheme} \label{III}

\subsection{Purely $\msb$ to $\msb$ matching}

In this section, we re-examine the one loop calculation of the matching between the $\msb$ renormalized quasi-PDF and the $\msb$ renormalized lightcone PDF following the steps of Ref. \cite{Izubuchi:2018srq}. We focus on the unpolarized isovector quark quasi-PDF in dimensional regularization with spacetime dimension $d=4-2\epsilon$. For this flavor non-singlet channel, mixing is not important so we dropped the flavour indices in the following discussion. The Dirac structure in the quark bilinear is chosen to be $\gamma^t$ which is typically used on the lattice to avoid mixing due to discretization induced chiral symmetry breaking \cite{Constantinou:2017sej,Green:2017xeu,Chen:2017mie}.
\begin{figure}[t]
    \centering
\begin{tikzpicture}
\begin{feynman}
\node (a) at (14.3,3.1);
\node (b) at (15.8,3.1);
\vertex (a1);
\vertex [above=of a1] (b1);
\vertex [above=of b1] (f11);
\vertex [right=of b1] (d1);
\vertex [right=of d1] (c1);
\vertex [above=of c1] (f21);
\vertex [below=of c1] (f31);
\vertex [right=of f31] (a2);
\vertex [above=of a2] (b2);
\vertex [above=of b2] (f12);
\vertex [right=of f12] (g2);
\vertex [right=of b2] (d2);
\vertex [right=of d2] (c2);
\vertex [above=of c2] (f22);
\vertex [below=of c2] (f32);
\vertex [right=of f32] (a3);
\vertex [above=of a3] (b3);
\vertex [above=of b3] (f13);
\vertex [right=of f13] (g3);
\vertex [right=of b3] (d3);
\vertex [right=of d3] (c3);
\vertex [above=of c3] (f23);
\vertex [below=of c3] (f33);
\vertex [right=of f33] (a4);
\vertex [above=of a4] (b4);
\vertex [above=of b4] (f14);
\vertex [right=of b4] (d4);
\vertex [right=of d4] (c4);
\vertex [above=of c4] (f24);
\vertex [below=of c4] (f34);
\diagram* {
(a1) -- [fermion, momentum={\(p\)}] (b1) -- [fermion, momentum={\(k\)}] (f11),
(b1) -- [gluon] (c1),
(f21) -- [fermion, momentum={\(k\)}] (c1),
(c1) -- [fermion, momentum={\(p\)}] (f31),
(f11) --[double=white,thick, edge label=\(z\)] (f21),
(a2) -- [fermion, momentum={\(p\)}](f12),
(g2) -- [gluon] (c2),
(f22) -- [fermion, momentum={\(k\)}] (c2),
(c2) -- [fermion, momentum={\(p\)}] (f32),
(f12) --[double=white,thick, edge label=\(z\)] (f22),
(a3) -- [fermion, momentum={\(p\)}] (b3) -- [fermion, momentum={\(k\)}] (f13),
(b3) -- [gluon] (g3),
(f23) -- [fermion, momentum={\(p\)}] (f33),
(f13) --[double=white,thick, edge label=\(z\)] (f23),
(a4) -- [fermion, momentum={\(p\)}] (f14),
(a) -- [gluon, half right] (b),
(f24) -- [fermion, momentum={\(p\)}] (f34),
(f14) --[double=white,thick, edge label=\(z\)] (f24)
};
\end{feynman}
\end{tikzpicture}
    \caption{Non-singlet quark quasi-PDF Feynman diagrams at one loop: the vertex(left), sail(middle two), and tadpole(right) diagrams.}
    \label{fig}
\end{figure}
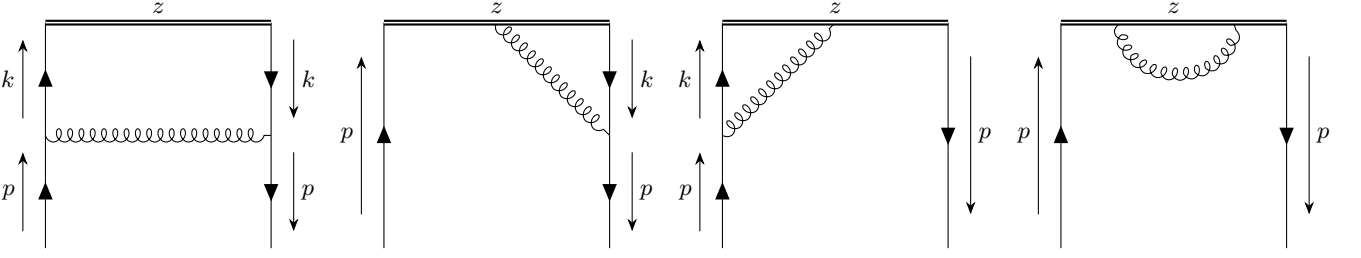
Here we choose the external quark state to be on-shell and massless. The UV and IR regulators are $\epsilon_{UV}>0$ and $\epsilon_{IR}<0$ respectively \cite{Izubuchi:2018srq}. We express the bare quasi-PDF of a quark state with a subscript $q$ as $\tilde{Q}^{B}_q$. In loop expansion,
\begin{align}
    \tilde{Q}^{B}_q(z,p^z,\tilde{\mu},\epsilon)&=\tilde{Q}^{(0)}(z,p^z)+\tilde{Q}^{(1)}(z,p^z,\tilde{\mu},\epsilon)+\mathcal{O}(\alpha_s^2),
\end{align}
with the tree level term
\begin{align}
    \tilde{Q}^{(0)}(z,p^z)=&e^{-ip^zz}.
\end{align}
In Fig.\ref{fig}, the one loop diagram of the quasi-PDF is shown. The diagrams are called the vertex(left), sail(middle two), and tadpole(right) diagrams. The one-loop contribution to the non-singlet bare quark quasi-PDF from each diagram is derived as below in the Feynman gauge \cite{Izubuchi:2018srq}:
\begin{align}
    \tilde{Q}_{vertex}^{(1)}(z,p^z,\epsilon,\tilde{\mu})=&\frac{\iota^\epsilon\tilde{\mu}^{2\epsilon}}{2p^t}\bar{u}(p)\int\frac{d^dk}{(2\pi)^d}(-igT^a\gamma^\mu)\frac{i\slashed{k}}{k^2}\gamma^t\frac{i\slashed{k}}{k^2}(-igT^a\gamma^\nu)\frac{-ig_{\mu\nu}}{(p-k)^2}u(p)e^{-ik^zz},\nonumber\\
    \tilde{Q}_{sail}^{(1)}(z,p^z,\epsilon,\tilde{\mu})=&\frac{\iota^\epsilon\tilde{\mu}^{2\epsilon}}{2p^t}\bar{u}(p)\int\frac{d^dk}{(2\pi)^d}\bigg\{(igT^a\gamma^t)\frac{1}{i(p^z-k^z)}\left(e^{-ip^zz}-e^{-ik^zz}\right)\delta^\mu_z\frac{i\slashed{k}}{k^2}(-igT^a\gamma^\nu)\frac{-ig_{\mu\nu}}{(p-k)^2},\nonumber\\
    &+(-igT^a\gamma^\nu)\frac{i\slashed{k}}{k^2}(igT^a\gamma^t)\frac{1}{i(p^z-k^z)}\left(e^{-ip^zz}-e^{-ik^zz}\right)\delta^\mu_z\frac{-ig_{\mu\nu}}{(p-k)^2}\bigg\}u(p),\nonumber\\
    \tilde{Q}_{tadpole}^{(1)}(z,p^z,\epsilon,\tilde{\mu})=&\frac{\iota^\epsilon\tilde{\mu}^{2\epsilon}}{2p^t}\bar{u}(p)\int\frac{d^dk}{(2\pi)^d}(-g^2)C_F\gamma^t\delta^\mu_z\delta^\nu_z\left(\frac{e^{-ip^zz}-e^{-ik^zz}}{(p^z-k^z)^2}-\frac{ze^{-ip^zz}}{i(p^z-k^z)}\right)\frac{-ig_{\mu\nu}}{(p-k)^2}u(p),\nonumber\\
    \tilde{Q}_{w.f.}^{(1)}(z,p^z,\epsilon,\tilde{\mu})=&-\frac{\alpha_sC_F}{4\pi}\left(\frac{1}{\epsilon_{UV}}-\frac{1}{\epsilon_{IR}}\right)e^{-ip^zz},\label{15}
\end{align}
where $\iota=e^{\gamma_E}/4\pi$ is for further convenience of $\msb$ subtraction, the $\tilde{Q}_{w.f.}^{(1)}$ stands for the one-loop quark wavefunction renormalization contribution and $\tilde{\mu}$ is the renormalization scale for the quasi-PDF. After the loop integral and the $\epsilon$ expansion, the total one loop contribution 
reproduces the result of Ref. \cite{Izubuchi:2018srq}:
\begin{align}
    \tilde{Q}^{(1)}(z,p^z,\tilde{\mu},\epsilon)=&\frac{\alpha_sC_F}{2\pi}\bigg\{\frac{3}{2}\left(\frac{1}{\epsilon_{UV}}+\ln{\frac{\tilde{\mu}^2z^2}{4e^{-2\gamma_E}}}+1\right)e^{-ip^zz}+\left(-\frac{1}{\epsilon_{IR}}-\ln{\frac{\tilde{\mu}^2z^2}{4e^{-2\gamma_E}}}-1\right)\nonumber\\
    &\times\left(\frac{3}{2}e^{-ip^zz}+\frac{1+ip^zz-e^{-ip^zz}-2ip^zz e^{-ip^zz}}{z^2p_z^2}-2e^{-ip^zz}\left(\Gamma(0,ip^zz)+\gamma_E+\ln(-ip^zz)
    \right)\right)\nonumber\\
    &+\frac{2(1-ip^zz-e^{-ip^zz})}{z^2p_z^2}+4ip^zz e^{-ip^zz}\pFq{3}{3}{1,1,1}{2,2,2}{ip^zz}\bigg\} ,
    \label{19}
\end{align}
where ${}_pF_q$ is a hypergeometric function. $\Gamma(a,b)$ is the incomplete gamma function defined as,
\begin{align}
    \Gamma(a,b)=\int^\infty_bt^{a-1}e^{-t}dt .
\end{align}
Now we can extract the counterterm from the bare quasi-PDF,
\begin{align}
    \tilde{Z}^{\msb}(\tilde{\mu},\epsilon)=&1+\frac{\alpha_sC_F}{2\pi}\frac{3}{2}\frac{1}{\epsilon_{UV}}+\mathcal{O}(\alpha_s^2).
    \label{21}
\end{align}

After renormalizing Eq.(\ref{19}) with Eq.(\ref{21}), the matching factor between the $\msb$ quasi-PDF to $\msb$ PDF can be obtained
by either performing a Fourier transform then using Eq.(\ref{9}) to obtain the kernel in momentum space, or using Eq.(\Ref{8})
to obtain the matching kernel in coordinate space then using Eq.(\Ref{10X}) to the momentum space. For $\msb$ to $\msb$ matching, these two approaches are equivalent. But for the hybrid to $\msb$ matching, the latter one is simpler. 

A more subtle issue is how to deal with the Fourier transform of $\ln z^2$, which appears after the $\epsilon$ expansion. This Fourier integral is not well defined. This raises the question whether one will obtain the same result for (a) computing the matching kernel directly using the momentum space Feynman rules and (b) computing the matching kernel in coordinate space then Fourier transform to the momentum space. Route (a) is to take the Fourier transform before $\epsilon$ expansion while route (b) follows the reversed order. So the question  reduces to whether the Fourier transformation and the $\epsilon$ expansion commute. Our answer is yes, like what was asserted in Ref.\cite{Izubuchi:2018srq}, but based on different arguments. We summarize our finding below. The details of our investigation can be found in Appendix \ref{A}.

The $\msb$ to $\msb$ matching kernel in momentum space  is \cite{Izubuchi:2018srq}
\begin{align}
    &C^{\msb}\left(\xi,\frac{\tilde{\mu}}{y P^z}\right)\nonumber\\
    =&\delta(1-\xi)+\frac{\alpha_sC_F}{2\pi}
    \begin{cases}
    \p{\frac{1+\xi^2}{1-\xi}\ln\frac{\xi}{\xi-1}+1+\frac{3}{2\xi}}{1}{\infty}{1}-\p{\frac{3}{2\xi}}{1}{\infty}{\infty}&,\xi>1\\
    \p{\frac{1+\xi^2}{1-\xi}\left(-\ln\frac{\tilde{\mu}^2}{y^2P_z^2}+\ln4\xi(1-\xi)\right)-\frac{\xi(1+\xi)}{1-\xi}}{0}{1}{1}&,0<\xi<1\\
    \p{-\frac{1+\xi^2}{1-\xi}\ln\frac{-\xi}{1-\xi}-1+\frac{3}{2(1-\xi)}}{-\infty}{0}{1}-\p{\frac{3}{2(1-\xi)}}{-\infty}{0}{-\infty}&,\xi<0
    \end{cases}\nonumber\\
    &+\frac{\alpha_sC_F}{2\pi}\left(\frac{3}{2}\ln\frac{\tilde{\mu}^2}{4y^2P^2_z}+\frac{5}{2}\right)\left(\delta(1-\xi)-\frac{1}{2}\left(\frac{1}{(1-\xi)^2}\delta^+\left(\frac{1}{1-\xi}\right)+\frac{1}{(\xi-1)^2}\delta^+\left(\frac{1}{\xi-1}\right)\right)\right)+\mathcal{O}(\alpha_s^2) \label{24}
\end{align}
where 
$\delta^+(1/x)$ is the delta function \New{(please see an example in Appendix \ref{A3})} with the argument being positive and
the plus function is defined as
\begin{equation}
    \p{f(x)}{a}{b}{c}\equiv  \int_a^bdx(f(x)-f(c)) , 
\end{equation}
\New{with
\begin{equation}
\int^a_bdx\p{f(x)}{a}{b}{\pm\infty}g(x)=\int^a_bdxf(x)(g(x)-g(\pm\infty))=\lim_{\beta\rightarrow0^\pm}\int^a_bdxf(x)(g(x)-g(1/\beta)) .
\end{equation}}
Quark number conservation is formally preserved because $\int d \xi C^{\msb} = 1$.

\Blue{In Ref. \cite{Izubuchi:2018srq}, same expression as Eq.(\ref{24}) is obtained if Fourier transform is performed before the $\epsilon$ expansion. However, a different expression, which does not preserve quark number, is obtained if $\epsilon$ expansion is performed first.  This is because the coordinate space correlator
$\tilde{Q}^{(1)}(z)$ in Eq.(\ref{19}) has a $\ln z^2$ dependence after the $\epsilon$ expansion. So it does not vanish as $z \to 0$.  
However, if $z=0$ is taken before the integration in Eq.(\ref{15}), then vector current conservation yields 
$\tilde{Q}^{(1)}(z=0)=0$. Therefore, the $z \to 0$ limit in Eq.(\ref{19}) is not continuous. 
} 

\Blue{At one loop, we follow the method of Ref. \cite{Izubuchi:2018srq} and recast $\ln z^2$ as a derivative of a power law in Eq.(\ref{A7}), which has a similar structure as resumming large logarithms and allows the Fourier transform being carried out. However, as shown in Eq.(\ref{A8}), 
there is an ambiguity in this approach which caused the non-commutativity of Fourier transform and $\epsilon$ expansion in Ref. \cite{Izubuchi:2018srq}. One can take the prescription of maintaining \New{the formal} quark number conservation to fix the ambiguity. By doing this, the commutativity of Fourier transform and $\epsilon$ expansion is also restored and the kernel yields the result of Eq.(\ref{24}).} 

%

\Blue{Ref. \cite{Izubuchi:2018srq} found, however, that the terms that caused the non-commutativity of Fourier transform and $\epsilon$ expansion 
did not contribute in the matching. Hence effectively these problems disappeared. In Appendix \ref{A3}, we re-examine how these terms, which are proportional to $\delta$ functions at infinite $|\xi|$, behave in the matching. We find that 
as long as the PDF's in the matching have finite net quark numbers, then these $\delta$ functions do not contribute in the matching. Ref. \cite{Izubuchi:2018srq} also reached a similar conclusion. However, their conclusion was based on the requirement that both the quark and antiquark numbers, instead of the net quark number, were finite in the PDF's of hadrons. This is not supported in global fits. However, despite of the defect, their conclusion is still correct. 
} 

\New{The fact that terms proportional to $\delta$ functions at infinite $|\xi|$ have no contribution in the matching has a profound implication. After dropping these terms, quark number is no longer conserved in Eq.(\ref{24}). This contradicts with the quark number conservation relation $\int d \xi C^{\msb} = 1$. The
contradiction
arises from integrating Eq.(\ref{A10}) over $\xi$. If $\beta \to 0^+$ limit is taken before the integration, as it should be, then the integral is zero. However, if the integration is carried out first, then the integral is finite. Therefore, we conclude that the $\msb$ to $\msb$ matching kernel in Eq.(\ref{24}) does not conserve quark number after all. Fortunately, this problem is fixed in the schemes of RI/MOM, ratio, and their corresponding hybrid versions such that quark number is conserved in these schemes.}

\subsection{Ratio and hybrid-ratio schemes}
\label{ratio}

\Blue{Now we move on to the ratio scheme to $\msb$ scheme matching, then work out the hybrid-ratio scheme to $\msb$ scheme matching.} 

Using Eq.(\ref{18}), the ratio scheme renormalization factor is 
\begin{align}
    \tilde{Z}^{\text{ratio}}(z,\tilde{\mu},\epsilon)=&\tilde{Q}^B_q(z, p^z=0,\epsilon)\nonumber\\
    =&1+\frac{\alpha_sC_F}{2\pi}\left(\frac{2}{3\epsilon_{UV}}+\frac{3}{2}\ln{\frac{\tilde{\mu}^2z^2}{4e^{-2\gamma_E}}}+\frac{5}{2}\right)+\mathcal{O}(\alpha_s^2) .
    \label{26}
\end{align}
Using Eqs.(\ref{10},\ref{21},\ref{26}), the conversion factor between the ratio scheme and the $\msb$ scheme becomes
\begin{align}
    Z^{\text{ratio}}_{\msb}\left(z,\tilde{\mu}\right)=1-\frac{\alpha_sC_F}{2\pi}\left(\frac{3}{2}\ln{\frac{\tilde{\mu}^2z^2}{4e^{-2\gamma_E}}}+\frac{5}{2}\right)+\mathcal{O}(\alpha_s^2) .\label{20}
\end{align}
Then from Eq.(\ref{12}) and the fact that $\mathcal{C}^{\msb}=\delta(1-\alpha)+\mathcal{O}(\alpha_s)$,  
the matching factor between the ratio to $\msb$ scheme in coordinate space is 
\begin{align}
    \mathcal{C}^{\text{ratio}}\left(\alpha,z^2\mu^2,\frac{\tilde{\mu}^2}{\mu^2}\right)
    =& \mathcal{C}^{\msb}(\alpha,z^2\mu^2)+\left(Z^{\text{ratio}}_{\msb}\left(z,\tilde{\mu}\right)-1\right)\delta(1-\alpha) +\mathcal{O}(\alpha_s^2) \nonumber \\
    =&\mathcal{C}^{\msb}(\alpha,z^2\mu^2)-\frac{\alpha_sC_F}{2\pi}\left(\frac{3}{2}\ln{\frac{\tilde{\mu}^2z^2}{4e^{-2\gamma_E}}}+\frac{5}{2}\right)\delta(1-\alpha)+\mathcal{O}(\alpha_s^2) .
    \label{28x}
\end{align}
Therefore, using \New{Eq.(\ref{10X})}, the ratio scheme to $\msb$ scheme matching factor in the momentum space is 
\begin{align}
    &C_{\gamma^t}^{\text{ratio}}\left(\xi,\frac{\tilde{\mu}}{y P^z}\right)\nonumber\\
    =&\delta(1-\xi)+\frac{\alpha_sC_F}{2\pi}
    \begin{cases}
    \p{\frac{1+\xi^2}{1-\xi}\ln\frac{\xi}{\xi-1}+1-\frac{3}{2(1-\xi)}}{1}{\infty}{1}&,\xi>1 ,\\
    \p{\frac{1+\xi^2}{1-\xi}\left(-\ln\frac{\tilde{\mu}^2}{y^2P_z^2}+\ln4\xi(1-\xi)-1\right)+1+\frac{3}{2(1-\xi)}}{0}{1}{1}&,0<\xi<1 ,\\
    \p{-\frac{1+\xi^2}{1-\xi}\ln\frac{-\xi}{1-\xi}-1+\frac{3}{2(1-\xi)}}{-\infty}{0}{1}&,\xi<0 .
    \end{cases}
    +\mathcal{O}(\alpha_s^2) , 
    \label{27}
\end{align}
where we have added the subscript $\gamma^t$ to mark it as an unpolarized case.
\Blue{This is identical to the result of Ref. \cite{Cichy:2018mum}. The $\delta$ functions at infinity $|\xi|$, which show up in $\mathcal{C}^{\msb}$ in Eq.(\ref{24}) are now cancelled. 
This is expected, because the $1/\epsilon_{UV}$ divergence is canceled in the ratio scheme, so are the associated $\ln z^2$ terms in short distance 
in the coordinate space and the $\delta$ functions at infinity $|\xi|$ in the momentum space. More explicitly, in Eq.(\ref{19}), the first $\ln z^2$ term associated with the $1/\epsilon_{UV}$ is canceled by the counterterm, while the second $\ln z^2$ term associated with the $1/\epsilon_{IR}$ has a vanishing prefactor as $z \to 0$. Therefore, without the $\ln z^2$ term at small $z$, the $z \to 0$ limit becomes smooth which in turn implies manifest quark number conservation.}


Despite this nice feature in the ratio scheme, in Eq.(\ref{20})
the scheme conversion factor $Z^{\text{ratio}}_{\msb}$ 
contains a $\ln{z^2\tilde{\mu}^2}$ term at one loop which becomes non-perturbative in IR (or large $z$). However, the conversion factor is a ratio of counterterms which should only have perturbative UV contributions. 
Ref. \cite{Ji:2020brr} argued that this was a drawback of the ratio scheme which could be remedied by the hybrid scheme to 
change the renormalization scheme at large $z$ to a Wilson line mass subtraction scheme. Then as shown in Eq.(\ref{14}), the corresponding conversion factor becomes 
\begin{align}
    Z^{\text{hybrid-ratio}}_{\msb}(z,z_s,\tilde{\mu})=&Z_{\msb}^{\text{ratio}}\left(z,\tilde{\mu}\right)\theta(z_s-\vert z\vert)+Z_{\msb}^{\text{ratio}}\left(z_s,\tilde{\mu}\right)\theta(\vert z\vert-z_s) .\label{30}
\end{align}
And the matching factor of the hybrid-ratio scheme in the coordinate space is
\begin{align}
    &\mathcal{C}^{\text{hybrid-ratio}}\left(\alpha,z^2\mu^2,z_s^2\mu^2,\frac{\tilde{\mu}^2}{\mu^2}\right) \nonumber\\
    =&\mathcal{C}^{\msb}(\alpha,z^2\mu^2)+\left(Z^{\text{hybrid-ratio}}_{\msb}\left(z,\tilde{\mu}\right)-1\right) \delta(1-\alpha)+\mathcal{O}(\alpha_s^2), \nonumber\\
   =&\mathcal{C}^{\text{ratio}}\left(\alpha,z^2\mu^2,\frac{\tilde{\mu}^2}{\mu^2}\right)+\left(Z^{\text{ratio}}_{\msb}\left(z_s,\tilde{\mu}\right)-Z^{\text{ratio}}_{\msb}\left(z,\tilde{\mu}\right)\right) \theta(\vert z\vert-z_s)\delta(1-\alpha)+\mathcal{O}(\alpha_s^2),
    \label{31X}
\end{align}
or
\begin{align}
    \mathcal{C}^{\text{hybrid-ratio}}\left(\alpha,z^2\mu^2,z_s^2\mu^2,\frac{\tilde{\mu}^2}{\mu^2}\right)
    =&\mathcal{C}^{\text{ratio}}\left(\alpha,z^2\mu^2,\frac{\tilde{\mu}^2}{\mu^2}\right)+\delta(1-\alpha)\frac{3\alpha_sC_F}{4\pi}\ln{\frac{z^2}{z_s^2}\theta(\vert z\vert-z_s)}+\mathcal{O}(\alpha_s^2).
    \label{31}
\end{align}
\Blue{Quark number conservation is manifest in this expression since, as mentioned above, $\mathcal{C}^{\text{ratio}}$ itself conserves quark number, and the second term vanishes as $z \to 0$ due to the $\theta$ function.
Then using Eq.(\ref{10X}) and the steps detailed in Appendix \ref{A4X},} the matching factor of the hybrid-ratio scheme to $\msb$ scheme in the momentum space is
\begin{align}
    C^{\text{hybrid-ratio}}_{\gamma^t}\left(\xi,y z_sP^z,\frac{\mu}{y P^z}\right)=&C^{\text{ratio}}_{\gamma^t}\left(\xi,\frac{\tilde{\mu}}{y P^z}\right)+\frac{3\alpha_sC_F}{4\pi}
    \p{-\frac{1}{\vert1-\xi\vert}+\frac{2\text{Si}((1-\xi) \vert y \vert
    z_sP^z)}{\pi(1-\xi)}}{-\infty}{\infty}{1}+\mathcal{O}(\alpha_s^2)\nonumber\\
    =&\delta(1-\xi)+\frac{\alpha_sC_F}{2\pi}
    \begin{cases}
    \p{\frac{1+\xi^2}{1-\xi}\ln\frac{\xi}{\xi-1}+1}{1}{\infty}{1}&,\xi>1\\
    \p{\frac{1+\xi^2}{1-\xi}\left(-\ln\frac{\tilde{\mu}^2}{y^2P_z^2}+\ln4\xi(1-\xi)-1\right)+1}{0}{1}{1}&,0<\xi<1\\
    \p{-\frac{1+\xi^2}{1-\xi}\ln\frac{-\xi}{1-\xi}-1}{-\infty}{0}{1}&,\xi<0
    \end{cases}\nonumber\\
    &+\frac{3\alpha_sC_F}{2\pi^2}\p{\frac{\text{Si}((1-\xi)
    \vert y \vert
    z_sP^z)}{(1-\xi)}}{-\infty}{\infty}{1}+\mathcal{O}(\alpha_s^2) , \label{28}
\end{align}
where we have added the $\gamma^t$ subscript for the unpolarized PDF and  
the plus function is for the $\xi$ variable only---it has no effect on the $|y|$ factor. 
$\text{Si}(x)$ is the sine integral defined as
\begin{equation}
    \text{Si}(x)\equiv\int_0^x\frac{\sin{t}}{t}dt .
\end{equation}
The one loop correction of the matching kernel in Eq.(\ref{28}) is written in plus functions. Hence quark number conservation is manifestly satisfied. This kernel in Eq.(\ref{28}), however, differs from the one of Ref. \cite{Ji:2020brr} in the last term
\Blue{although both derivations agree on Eqs.(\ref{31}) and (\ref{10X}).}
Ref. \cite{Ji:2020brr} has
\begin{equation}
    \frac{3\alpha_sC_F}{2\pi}\frac{\text{Si}((1-\xi)y z_sP^z)}{\pi(1-\xi)} 
    \label{33}
\end{equation}
instead. \Blue{In addition to the difference between $y$ and $|y|$ in the argument of the Si function, quark number conservation is violated in the expression of (\ref{33}) but is preserved in (\ref{28}).} 

\begin{figure}[t]
    \includegraphics[width=0.7 \textwidth]{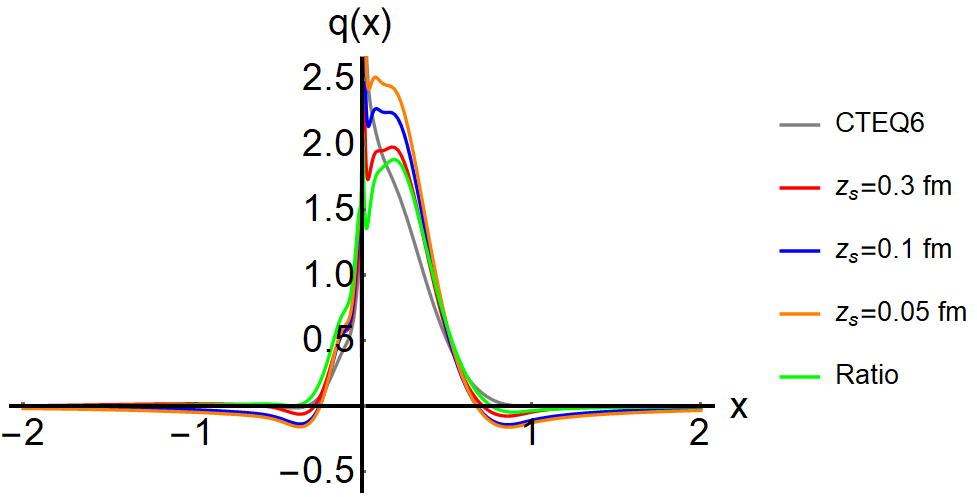}
    \includegraphics[width=0.49 \textwidth]{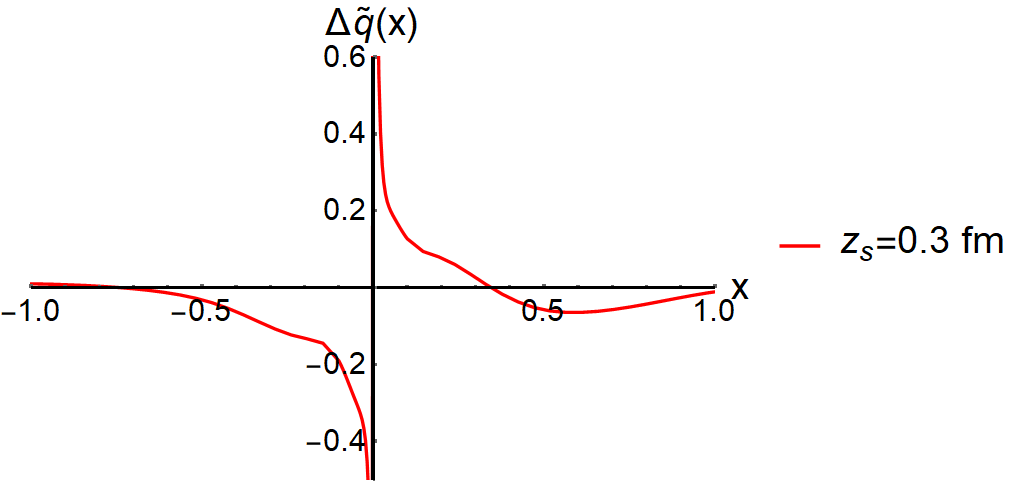}
    \includegraphics[width=0.49 \textwidth]{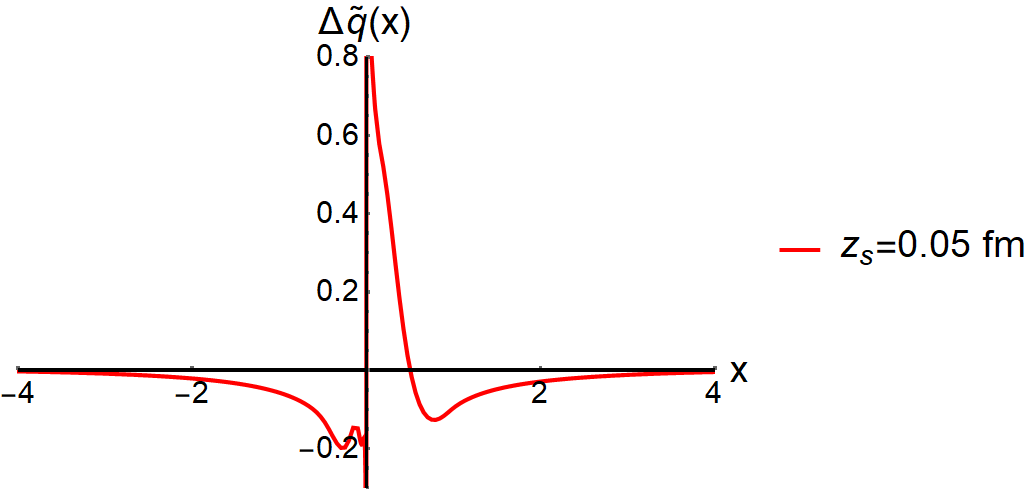}
    \caption{ Upper panel: unpolarized isovector proton quasi-PDFs in the ratio scheme and the hybrid-ratio scheme with varies $z_s$ values computed with the one loop matching formula. The inputs are the \New{CTEQ6} global fit  \cite{Pumplin:2002vw}, $\alpha_s=0.283$, $P_z=1.5$ GeV, and $\mu=3$ GeV. The hybrid-ratio curves coincide with the hybrid-RI/MOM curves with $p_z^R=0$. Lower panel: the difference between the $z_s=0.3(0.05)$ fm and ratio curve is shown in the left(right) figure. The area of the curve is zero by quark number conservation. }
    \label{f}
\end{figure}

In the upper panel of Fig. \ref{f}, the matching formula of Eq. (\ref{9}) and the one-loop matching factors derived in Eq. (\ref{27}) and Eq. (\ref{28}) are used to yield the corresponding quasi-PDFs in the ratio scheme and the hybrid ratio scheme. The input PDF 
is from the global fit of the CTEQ-JLab Collaboration(CJ12)\cite{Owens:2012bv} and the parameters used are $\alpha_s=0.283$, $P_z=1.5$ GeV, and $\mu=3$ GeV. Different values of $z_s$ are shown for the hybrid-ratio scheme. Formally, the ratio scheme is corresponding to the $z_s \to \infty$ limit of the hybrid-ratio scheme. However, \New{$a \ll z_s \lesssim 0.3$ fm, with $a$ the lattice spacing,
is recommended to avoid large discretization errors and higher twist contaminations \cite{Ji:2020brr}.} We see the $z_s = 0.3$ fm curve is quite close to the ratio scheme curve already. Reducing $z_s$ tends to increase the small and positive $x$ part while making the larger $|x|$ part more negative. In the lower panel, the difference between the $z_s=0.3(0.05)$ fm and ratio curve is shown in the left(right) figure. The area of the curve is zero by quark number conservation.
However, larger range of $x$ is needed for smaller $z_s$ for the integration. (Note that the ranges of $x$ for the two figures are different.)


Analogous to the unpolarized case in Eq.(\ref{28}), the matching factors of the hybrid-ratio scheme to $\msb$ scheme in the momentum space for helicity and transversity cases are,
\begin{align}
&\New{g_A} C^{\text{hybrid-ratio}}_{\gamma^z\gamma_5}\left(\xi,y z_sP^z,\frac{\mu}{y P^z}\right)\nonumber\\
=&\delta(1-\xi)+
\frac{\alpha_sC_F}{2\pi}\begin{cases}
    \p{\frac{1+\xi^2}{1-\xi}\ln\frac{\xi}{\xi-1}+1}{1}{\infty}{1}&,\xi>1\\
    \p{\frac{1+\xi^2}{1-\xi}\left(-\ln\frac{\tilde{\mu}^2}{y^2P_z^2}+\ln4\xi(1-\xi)\right)+\frac{2-5\xi+\xi^2}{1-\xi}}{0}{1}{1}&,0<\xi<1\\
    \p{-\frac{1+\xi^2}{1-\xi}\ln\frac{-\xi}{1-\xi}-1}{-\infty}{0}{1}&,\xi<0
    \end{cases}\nonumber\\
    &+\frac{3\alpha_sC_F}{2\pi^2}\p{\frac{\text{Si}((1-\xi)
    \vert y \vert
    z_sP^z)}{(1-\xi)}}{-\infty}{\infty}{1}+\mathcal{O}(\alpha_s^2) ,
    \label{36}
\end{align}
and
\begin{align}
&\New{g_T} C^{\text{hybrid-ratio}}_{\gamma^z\gamma^\perp}\left(\xi,y z_sP^z,\frac{\mu}{y P^z}\right)\nonumber\\
=&\delta(1-\xi)+\frac{\alpha_sC_F}{2\pi}\begin{cases}
    \p{\frac{2\xi}{1-\xi}\ln\frac{\xi}{\xi-1}}{1}{\infty}{1}&,\xi>1\\
    \p{\frac{2\xi}{1-\xi}\left(-\ln\frac{\tilde{\mu}^2}{y^2P_z^2}+\ln4\xi(1-\xi)-1\right)}{0}{1}{1}&,0<\xi<1\\
    \p{-\frac{2\xi}{1-\xi}\ln\frac{\xi}{\xi-1}}{-\infty}{0}{1}&,\xi<0
    \end{cases}\nonumber\\
    &+\frac{2\alpha_sC_F}{\pi^2}\p{\frac{\text{Si}((1-\xi)
    \vert y \vert
    z_sP^z)}{(1-\xi)}}{-\infty}{\infty}{1}+\mathcal{O}(\alpha_s^2) ,
    \label{37}
\end{align}
\New{with $g_A \simeq 1.275$ and $g_T = 0.99(4)$ \cite{Gupta:2018qil}.}

\New{The above hybrid-ratio results are computed with on-shell external quark states with dimensional regularization. The gauge invariant IR regulator $\epsilon_{IR}$ is used such that the results are gauge invariant. 
We have verified that these
hybrid-ratio kernels coincide with the hybrid-RI/MOM ones of Sec. \ref{h-RM}
in the limits of $p_R^z \to 0$ and $\mu_R \ll p^z$ (the order of taking the limits does not matter). We have also verified that the gauge dependence in the hybrid-RI/MOM kernels 
due to the off-shell quark IR regulators disappears after the IR singularities are cancelled between PDF's and quasi-PDF's.}

\subsection{Hybrid-RI/MOM scheme}
\label{h-RM}

The momentum space matching kernel between the RI/MOM quasi-PDF and $\msb$ PDF in one loop is 
\begin{align}
    C^{\text{RI/MOM}}_\Gamma(\xi,y,P^z,\tilde{\mu},p^z_R,\mu_R)=\delta(1-\xi)+C_\Gamma^{B}(\xi,y,P^z,\tilde{\mu}) +C^{\text{RI/MOM}}_{\text{CT},\Gamma}(\xi,y,P^z,\tilde{\mu},p^z_R,\mu_R) +\mathcal{O}(\alpha_s^2) ,
    \label{39}
\end{align}
with $\Gamma=\gamma^t,\,\gamma^z\gamma_5,\,\gamma^z\gamma^\perp$ for the unpolarized, helicity, and transversity PDF's respectively. These functions can be found in Refs. \cite{Liu:2018hxv,LatticeParton:2018gjr}. They are derived in Landau gauge which is typically employed on the lattice and minimal projection described in Ref \cite{LatticeParton:2018gjr}. The part associated with the bare quasi-PDF is
\begin{align}
&C_{\gamma^t}^{B}(\xi,y,P^z,\tilde{\mu})=\frac{\alpha_sC_F}{2\pi}\begin{cases}
    \p{\frac{1+\xi^2}{1-\xi}\ln\frac{\xi}{\xi-1}+1}{1}{\infty}{1}&,\xi>1\\
    \p{\frac{1+\xi^2}{1-\xi}\left(-\ln\frac{\tilde{\mu}^2}{y^2P_z^2}+\ln4\xi(1-\xi)-1\right)+1}{0}{1}{1}&,0<\xi<1\\
    \p{-\frac{1+\xi^2}{1-\xi}\ln\frac{-\xi}{1-\xi}-1}{-\infty}{0}{1}&,\xi<0
    \end{cases}\nonumber\\
&C_{\gamma^z\gamma_5}^{B}(\xi,y,P^z,\tilde{\mu})=\frac{\alpha_sC_F}{2\pi}\begin{cases}
    \p{\frac{1+\xi^2}{1-\xi}\ln\frac{\xi}{\xi-1}+1}{1}{\infty}{1}&,\xi>1\\
    \p{\frac{1+\xi^2}{1-\xi}\left(-\ln\frac{\tilde{\mu}^2}{y^2P_z^2}+\ln4\xi(1-\xi)\right)+\frac{2-5\xi+\xi^2}{1-\xi}}{0}{1}{1}&,0<\xi<1\\
    \p{-\frac{1+\xi^2}{1-\xi}\ln\frac{-\xi}{1-\xi}-1}{-\infty}{0}{1}&,\xi<0
    \end{cases}\nonumber\\
&\New{C_{\gamma^z\gamma^\perp}^{B}(\xi,y,P^z,\tilde{\mu})=\frac{\alpha_sC_F}{2\pi}\begin{cases}
    \p{\frac{2\xi}{1-\xi}\ln\frac{\xi}{\xi-1}-\frac{1}{2(1-\xi)}}{1}{\infty}{1}+\frac{1}{2(1-\xi)}&,\xi>1\\
    \p{\frac{2\xi}{1-\xi}\left(-\ln\frac{\tilde{\mu}^2}{y^2P_z^2}+\ln4\xi(1-\xi)-1\right)+\frac{1}{2(1-\xi)}}{0}{1}{1}-\frac{1}{2(1-\xi)}&,0<\xi<1\\
    \p{-\frac{2\xi}{1-\xi}\ln\frac{\xi}{\xi-1}+\frac{1}{2(1-\xi)}}{-\infty}{0}{1}-\frac{1}{2(1-\xi)}&,\xi<0
    \end{cases}}
    \label{40}
\end{align}
The counterterms can be expressed as
\begin{align}
    C^{\text{RI/MOM}}_{\text{CT},\Gamma}(\xi,y,P^z,\tilde{\mu},p^z_R,\mu_R)=-\left\vert\frac{p^z}{p^z_R}\right\vert h_\Gamma\left(\frac{p^z}{p^z_R}(\xi-1)+1,r\right),
\end{align}
with $p^z=yP^z$, $r=\mu_R^2/{p^z_R}^2 \New{-i 0^+}$. 
\begin{align}
&h_{\gamma^t}\left(x,r\right)=\frac{\alpha_sC_F}{2\pi}\begin{cases}
    \p{\frac{-3r^2+13rx-8x^2-10rx+8x^3}{2(r-1)(x-1)(r-4x+4x^2)}+\frac{-3r+8x-rx+4x^2}{2(r-1)^{3/2}(x-1)}\tan^{-1}\frac{\sqrt{r-1}}{2x-1}}{1}{\infty}{1}&,x>1\\
    \p{\frac{-3r+7x-4x^2}{2(r-1)(1-x)}+\frac{3r-8x+rx+4x^2}{2(r-1)^{3/2}(1-x)}\tan^{-1}\sqrt{r-1}}{0}{1}{1}&,0<x<1\\
    \p{-\frac{-3r^2+13rx-8x^2-10rx+8x^3}{2(r-1)(x-1)(r-4x+4x^2)}-\frac{-3r+8x-rx+4x^2}{2(r-1)^{3/2}(x-1)}\tan^{-1}\frac{\sqrt{r-1}}{2x-1}}{-\infty}{-1}{1}&,x<0\\
    \end{cases}\nonumber\\
&h_{\gamma^z\gamma_5}\left(x,r\right)=\frac{\alpha_sC_F}{2\pi}\begin{cases}
\p{\frac{3r-(1-2x)^2}{2(r-1)(1-x)}-\frac{4x^2(2-3r+2x+4rx-12x^2+8x^3)}{(r-1)(r-4x+4x^2)^2}+\frac{2-3r+2x^2}{(r-1)^{3/2}(x-1)}\tan^{-1}\frac{\sqrt{r-1}}{2x-1}}{1}{\infty}{1}&,x>1\\
\p{\frac{1-3r+4x^2}{2(r-1)(1-x)}+\frac{-2+3r-2x^2}{(r-1)^{3/2}(1-x)}\tan^{-1}\sqrt{r-1}}{0}{1}{1}&,0<x<1\\
\p{-\frac{3r-(1-2x)^2}{2(r-1)(1-x)}+\frac{4x^2(2-3r+2x+4rx-12x^2+8x^3)}{(r-1)(r-4x+4x^2)^2}-\frac{2-3r+2x^2}{(r-1)^{3/2}(x-1)}\tan^{-1}\frac{\sqrt{r-1}}{2x-1}}{-\infty}{-1}{1}&,x<0\\
\end{cases}\nonumber\\
&\New{h_{\gamma^z\gamma^\perp}\left(x,r\right)=\frac{\alpha_sC_F}{2\pi}\begin{cases}
\p{\frac{3 r+8 (x-1) x}{2(1-x)(r-4x+4x^2)}+\frac{1+x}{\sqrt{r-1}(1-x)}\tan^{-1}\frac{\sqrt{r-1}}{2x-1}}{1}{\infty}{1}\\
+\frac{r(1-2x)}{(r-1)(r-4x+4x^2)}+\frac{1}{(r-1)^{\frac{3}{2}}}\tan^{-1}\frac{\sqrt{r-1}}{2x-1}&,x>1\\
\p{-\frac{3}{2(1-x)}+\frac{1+x}{\sqrt{r-1}(1-x)}\tan^{-1}\sqrt{r-1}}{0}{1}{1}\\
-\frac{1}{r-1}+\frac{1}{(r-1)^{\frac{3}{2}}}\tan^{-1}\sqrt{r-1}&,0<x<1\\
\p{-\frac{3 r+8 (x-1) x}{2(1-x)(r-4x+4x^2)}-\frac{1+x}{\sqrt{r-1}(1-x)}\tan^{-1}\frac{\sqrt{r-1}}{2x-1}}{-\infty}{-1}{1}\\
-\frac{r(1-2x)}{(r-1)(r-4x+4x^2)}-\frac{1}{(r-1)^{\frac{3}{2}}}\tan^{-1}\frac{\sqrt{r-1}}{2x-1}&,x<0\\
\end{cases}}
\end{align}

The analysis is analogous to Eq.(\ref{31X}), but $C^B$ in Eq.(\ref{39}) is not $C^{\msb}$. It is bare in the quasi-PDF side whose $\epsilon$ pole in the virtual diagram would be cancelled by that in $C^{\text{RI/MOM}}_{\text{CT}}$. Hence to hybridize this RI/MOM scheme result, we need to Fourier transform the counetrterm $C^{\text{RI/MOM}}_{\text{CT}}$ and modify it in the coordinate space then Fourier transform it back to the momentum space. 

Now we rewrite $C^{\text{RI/MOM}}_{\text{CT}}$ as
\begin{align}
    C^{\text{RI/MOM}}_{\text{CT},\Gamma}(\xi,y,P^z,\tilde{\mu},p^z_R,\mu_R)    =&-\p{\left\vert\frac{p^z}{p^z_R}\right\vert h_{\Gamma,p}\left(\frac{p^z}{p^z_R}(\xi-1)+1,r\right)}{-\infty}{\infty}{1}-\left\vert\frac{p^z}{p^z_R}\right\vert h'_{\Gamma}\left(\frac{p^z}{p^z_R}(\xi-1)+1,r\right)
\end{align}
with $h_{\Gamma,p}$ the plus function part and $h'_{\Gamma}$ the non-plus function part. Then analogous to Eq.(\ref{31X}), through Eq.(\ref{10X}),  
\begin{align}
\tilde{Z}^{\text{RI/MOM}}(z,p^z_R,\epsilon,\mu_R)-1 
=&\int d\xi e^{-i(\xi-1) yP^zz}C^{\text{RI/MOM}}_{\text{CT},\Gamma}(\xi,y,P^z,\tilde{\mu},p^z_R,\mu_R)\nonumber\\
=&-\int dx e^{i(p^z_R(1-x))z}(h_{\Gamma,p}(x,r)+h'_{\Gamma}\left(x,r\right))+\int dx h_{\Gamma,p}(x,r) .
\label{43}
\end{align}
Using this we can construct the renormalization factor of hybrid-RI/MOM in coordinate space
\begin{align}
Z^{\text{hybrid-RI/MOM}}(z)-1 =&
(\tilde{Z}^{\text{RI/MOM}}(z)-1)\theta(z_s-|z|)+(\tilde{Z}^{\text{RI/MOM}}(z_s)-1)\theta(|z|-z_s)\nonumber \\
=&(\tilde{Z}^{\text{RI/MOM}}(z)-\tilde{Z}^{\text{RI/MOM}}(z_s))\theta(z_s-|z|)+(\tilde{Z}^{\text{RI/MOM}}(z_s)-1) ,
\label{B10}
\end{align}
which can be transformed back to momentum space using Eq.(\Ref{10X}) again or the reverse of Eq.(\Ref{43}):
\begin{align}
    C^{\text{hybrid-RI/MOM}}_{\text{CT},\Gamma}(\xi,y)
    =\int\frac{P^zdz}{2\pi}e^{i(\xi-1)P^zz}\left(Z^{\text{hybrid-RI/MOM}}\left(\frac{z}{y}\right)-1\right) .
\end{align}
This yields
\begin{align}
    &C^{\text{hybrid-RI/MOM}}_{\text{CT},\Gamma}(\xi,y)\nonumber\\
    =&\p{\int^\infty_{-\infty}dx\left(\frac{e^{i(1-x)p^z_Rz_s}\sin((\xi-1)z_s|y|P^z)}{\pi(\xi-1)}-\frac{|y|P^z\sin((yP^z(\xi-1)-p^z_R(x-1))z_s)}{\pi(yP^z(\xi-1)-p^z_R(x-1))}\right)h_{\Gamma,p}(x,r)}{-\infty}{\infty}{1}\nonumber\\
    &+\p{\int^\infty_{-\infty}dx\left(\frac{e^{i(1-x)p^z_Rz_s}\sin((\xi-1)z_s|y|P^z)}{\pi(\xi-1)}\right)h'_\Gamma(x,r)}{-\infty}{\infty}{1}\nonumber\\
    &-\int^\infty_{-\infty}dx\left(\frac{|y|P^z\sin((yP^z(\xi-1)-p^z_R(x-1))z_s)}{\pi(yP^z(\xi-1)-p^z_R(x-1))}\right)h'_\Gamma(x,r) .
\end{align}
Finally, the momentum space matching kernel between the hybrid-RI/MOM quasi-PDF and $\msb$ PDF at one loop is
\begin{align}
    C^{\text{hybrid-RI/MOM}}_\Gamma(\xi,y,P^z,\tilde{\mu},p^z_R,\mu_R,z_s)=\delta(1-\xi)+C_\Gamma^{B}(\xi,y,P^z,\tilde{\mu}) +C^{\text{hybrid-RI/MOM}}_{\text{CT},\Gamma}(\xi,y,P^z,\tilde{\mu},p^z_R,\mu_R,z_s) +\mathcal{O}(\alpha_s^2) .
    \label{47}
\end{align}
There are some interesting limits for this result:

(1) When 
\New{$p^z_R \to 0$ and $\mu_R \ll p^z =yP^z$ (the order does not matter)}, the result is reduced to the hybrid-ratio result shown in Eqs. (\ref{28},\ref{36},\ref{37}):
\begin{align}
    C^{\text{hybrid-RI/MOM}}_\Gamma(\xi,y,P^z,\mu,p^z_R = 0,\mu_R,z_s)\bigg\vert_{\mu_R z_s \ll 1}=g_{\Gamma}C^{\text{hybrid-ratio}}_{\Gamma}\left(\xi,y z_sP^z,\frac{\mu}{y P^z}\right) ,
\end{align}
where $g_{\Gamma}$ defined in Eq.(\ref{18}) is the charge of the PDF. $g_{\gamma^t}=1$, $g_{\gamma^z\gamma_5}=g_A \simeq 1.275$ and $g_{\gamma^z\gamma^{\perp}}=g_T = 0.99(4)$ \cite{Gupta:2018qil}. \New{As expected, the gauge dependence disappears in this limit.}

(2) When $z_s \to \infty$, the RI/MOM result is recovered:
\begin{align}
    C^{\text{hybrid-RI/MOM}}_\Gamma(\xi,y,P^z,\mu,p^z_R,\mu_R,z_s)\bigg\vert_{z_s \to \infty}=C^{\text{RI/MOM}}_\Gamma(\xi,y,P^z,\tilde{\mu},p^z_R,\mu_R) .
\end{align}

(3) When $z_s \to 0$, the self renormalization scheme result is recovered:
\begin{align}
    C^{\text{hybrid-RI/MOM}}_\Gamma(\xi,y,P^z,\mu,p^z_R,\mu_R,z_s=0)
    =C^{\text{self}}_\Gamma(\xi,y,P^z,\tilde{\mu},p^z_R,\mu_R)=C^{\msb}_\Gamma\left(\xi,\frac{\tilde{\mu}}{y P^z}\right) .
    \label{50}
\end{align}
Note that a non-zero $z_s$ is already assumed in Eq.(\ref{47}) such that the $1/\epsilon$ poles in virtual diagrams in $C^B$ will be cancelled by the the counterterm $C^{\text{RI/MOM}}_{\text{CT}}$. For $z_s=0$, we can use Eq.(\ref{31X}) and
$Z^{\text{hybrid-ratio}}_{\msb} =1$. The result is $C^{\msb}$. However, $C^{\msb}(\xi)\propto 1/|\xi|$ as $|\xi| \to \infty$. That means the plus function is not well defined since the prefactor in front of $\delta(\xi-1)$ diverges. 
This suggests that one cannot use self renormalization to all range of $z$ and some modification of the scheme at small $z$ is needed.  
\Red{Indeed Ref.~\cite{LatticePartonCollaborationLPC:2021xdx} has realized this and fits the self renormalized matrix element to the $\msb$ one only down to a perturbative $z\gg a$. Also, a modified self renormalization or hybrid-ratio like scheme has been applied to supplement the self renormalization with short distance part renormalized in the ratio scheme~\cite{Hua:2022kcm}. This scheme is implemented to meson DA computations with the coordinate space matching kernel derived as well.}
\end{widetext}

\section{Conclusion} \label{V}

We have calculated the matching kernels for the unpolarized, helicity, and transversity isovector parton distribution functions and skewless generalized parton distributions of all hadrons in the hybrid-RI/MOM scheme. This result is connected to lots of special cases. For example, when $z_s \to \infty$, by design, the non-hybrid scheme is recovered. When $z_s \to 0$, the self renormalization scheme is obtained. Our analysis suggests that one cannot use self renormalization to all range of $z$ and some modification
of the scheme at small $z$ is needed.
When the parameters $p_z^R=0$ and $\mu^R z_s  \ll 1$, the hybrid-RI/MOM scheme coincides with the hybrid-ratio scheme times the charge of the PDF. We have also discussed the subtlety related to the commutativity of Fourier transform and $\epsilon$ expansion in the $\msb$ scheme. 


\begin{widetext}

\section*{Acknowledgement}
This work is partly supported by the Ministry of Science and
Technology, Taiwan, under Grant No. 108- 2112-M-002-
003-MY3 and the Kenda Foundation.

\appendix
\section{Fourier transform of singular functions} \label{A}
We demonstrate the equivalence of the one-step (i.e. performing the kernel computation in momentum space) and two-step (i.e. performing the kernel computation in coordinate space first then Fourier transforming it to the momentum space) matching in this Appendix. 

We need to show that two operations, Fourier transformed and $\epsilon$ expansion, commute when acting on 
the following function  
\begin{equation}
    f(z)=(\vert z\vert\mu)^{2\epsilon}\Gamma(-\epsilon) 
\end{equation}
arising from the integrals in Eq. (\ref{15}) \cite{Ji:2017rah}. The renormalization scale $\mu$ is added to keep $f(z)$ dimensionless for non-zero $\epsilon$. 

\subsection{Performing the Fourier transform first}\label{A1}
The Fourier transform of $f(z)$ yields
\begin{align}
    \tilde{f}(x)=&\int\frac{d\zeta}{2\pi}e^{ix\zeta}f(z)=\left(\frac{2\mu}{p^z}\right)^{2\epsilon}\frac{\Gamma\left(\epsilon+\frac{1}{2}\right)}{\sqrt{\pi}}\frac{1}{\vert x\vert^{1+2\epsilon}} \ ,\label{A2}
\end{align}
with $\zeta = z p^z$.
We can further $\epsilon$ expand the factor $1/|x|^{1+2\epsilon}$ by multiplying it by a test function $g(x)$ then performing an integration \cite{Fleming:2007xt}: 
\begin{align}
    \int^1_0dx\frac{1}{x^{1+2\epsilon}}g(x)=&\int^1_0dx\frac{1}{x^{1+2\epsilon}}g(0)+\int^1_0dx\frac{1}{x^{1+2\epsilon}}(g(x)-g(0))\nonumber\\
    =&\int^1_0dx\left(-\frac{1}{2\epsilon_{IR}}\delta(x)+\p{\frac{1}{x}}{0}{1}{0}\right)g(x)+\mathcal{O}(\epsilon),\\
    \int^\infty_1dx\frac{1}{x^{1+2\epsilon}}g(x)=&\int^\infty_1dx\frac{1}{x^{1+2\epsilon}}g(\infty)+\int^\infty_1dx\frac{1}{x^{1+2\epsilon}}(g(x)-g(\infty))\nonumber\\
    =&\int^\infty_1dx\left(\frac{1}{2\epsilon_{UV}}\frac{1}{x^2}\delta^+\left(\frac{1}{x}\right)+\p{\frac{1}{x}}{1}{\infty}{\infty}\right)g(x)+\mathcal{O}(\epsilon).
\end{align}
Putting them together, we have
\begin{align}
    \frac{\theta(x)}{x^{1+2\epsilon}}=&-\frac{1}{2\epsilon_{IR}}\delta(x)+\frac{1}{2\epsilon_{UV}}\frac{1}{x^2}\delta^+\left(\frac{1}{x}\right)+\p{\frac{1}{x}}{0}{1}{0}+\p{\frac{1}{x}}{1}{\infty}{\infty}+\mathcal{O}(\epsilon) .
\end{align}
Therefore, the $\epsilon$ expansion of Eq.(\ref{A2}) becomes
\begin{align}
    \tilde{f}(x)=&\left(\frac{2\mu}{p^z}\right)^{2\epsilon}\frac{\Gamma\left(\epsilon+\frac{1}{2}\right)}{\sqrt{\pi}}\frac{\theta(x)-\theta(-x)}{x^{1+2\epsilon}}\nonumber\\
    =&\left\{1+\epsilon\left(\ln\left(\frac{\mu^2}{p_z^2}\right)-\gamma_E\right)+\mathcal{O}(\epsilon^2)\right\}\bigg\{-\frac{1}{\epsilon_{IR}}\delta(x)+\frac{1}{\epsilon_{UV}}\frac{1}{2}\left(\frac{1}{x^2}\delta^+\left(\frac{1}{x}\right)+\frac{1}{(-x)^2}\delta^+\left(-\frac{1}{x}\right)\right)\nonumber\\
    &+\p{\frac{1}{x}}{0}{1}{0}+\p{\frac{1}{x}}{1}{\infty}{\infty}+\p{-\frac{1}{x}}{-1}{0}{0}+\p{-\frac{1}{x}}{-\infty}{-1}{-\infty}+\mathcal{O}(\epsilon)\bigg\}\nonumber\\
    =&-\frac{1}{\epsilon_{IR}}\delta(x)+\frac{1}{\epsilon_{UV}}\frac{1}{2}\left(\frac{1}{x^2}\delta^+\left(\frac{1}{x}\right)+\frac{1}{(-x)^2}\delta^+\left(-\frac{1}{x}\right)\right)\nonumber\\
    &+\left[\gamma_E-\ln\left(\frac{\mu^2}{p_z^2}\right)\right]\left[\delta(x)-\frac{1}{2}\left(\frac{1}{x^2}\delta^+\left(\frac{1}{x}\right)+\frac{1}{(-x)^2}\delta^+\left(-\frac{1}{x}\right)\right)\right]\nonumber\\
    &+\p{\frac{1}{x}}{0}{1}{0}+\p{\frac{1}{x}}{1}{\infty}{\infty}+\p{-\frac{1}{x}}{-1}{0}{0}+\p{-\frac{1}{x}}{-\infty}{-1}{-\infty}+\mathcal{O}(\epsilon)\ .
    \label{A4}
\end{align}
In the $\msb$ to $\msb$ matching kernel shown in Eq.(\ref{24}), neither $\epsilon_{UV}$ nor $\epsilon_{IR}$ appears because 
the $\epsilon_{UV}$ dependent UV divergent term is removed by renormalization, while the $\epsilon_{IR}$ dependent IR divergent term should not contribute to the matching kernel since the IR divergence of the PDF and the quasi-PDF cancel in the matching kernel.
As a result, we obtain the following contribution in the matching kernel,
\begin{align}
    \tilde{f}^C(x)=&\left[\gamma_E-\ln\left(\frac{\mu^2}{p_z^2}\right)\right]\left[\delta(x)-\frac{1}{2}\left(\frac{1}{x^2}\delta^+\left(\frac{1}{x}\right)+\frac{1}{(-x)^2}\delta^+\left(-\frac{1}{x}\right)\right)\right]\nonumber\\
    &+\p{\frac{1}{x}}{0}{1}{0}+\p{\frac{1}{x}}{1}{\infty}{\infty}+\p{-\frac{1}{x}}{-1}{0}{0}+\p{-\frac{1}{x}}{-\infty}{-1}{-\infty}+\mathcal{O}(\epsilon)\label{A5}
\end{align}
\subsection{Performing the $\epsilon$ expansion first}
After the $\epsilon$ expansion,
\begin{align}
    f(z)=&-\frac{1}{\epsilon_{IR}}+(-\gamma_E-\ln z^2\mu^2)+\mathcal{O}(\epsilon) ,
\end{align}
where the $\epsilon_{IR}$ pole is canceled by the PDF IR singularity in the coordinate space matching kernel. %
The Fourier transform of $\ln{z^2}$ 
\Blue{does not converge. Hence strictly speaking, performing the  $\epsilon$ expansion before Fourier transform is not well defined. Nevertheless, it was rewritten as the derivative of a power and computed in Ref.}
 \cite{Izubuchi:2018srq}
\begin{align}
\label{A7}
    -\int\frac{dzp^z}{2\pi}e^{ixzp^z}\ln(z^2 \mu^2 e^{\gamma_E})=&-\left[\frac{d}{d\eta}\int\frac{dzp^z}{2\pi}e^{ixzp^z}\left(z^2 \mu^2 e^{\gamma_E}\right)^\eta\right]\Bigg\vert_{\eta=0}\nonumber\\
    =&-\left[\frac{d}{d\eta}(\frac{\mu^2 e^{\gamma_E}}{p_z^2})^{\eta}\frac{4^\eta}{ \Gamma(-\eta)}\frac{\Gamma(\eta+1/2)}{\sqrt{\pi}}\frac{1}{\vert x\vert^{1+2\eta}}\right]\Bigg\vert_{\eta=0}\nonumber\\
    =& \tilde{f}^C(x) .
\end{align}
\Blue{Hence Eqs. (\ref{A5}) and (\ref{A7}) coincide, indicating that Fourier transformation and $\epsilon$ expansion commute. However, writing $\ln{z^2}$ as a power's derivative is similar to undoing the $\epsilon$ expansion. Hence the agreement is perhaps not a strict test of the commutativity of the two operations.}

It is worth noting that Ref. \cite{Izubuchi:2018srq} did not obtain the same result for the integral of Eq.(\ref{A7}). Hence the commutativity of Fourier transformation and $\epsilon$ expansion was not obtained in that work. This is due to the following ambiguity in the integral. By separating the logarithm into two terms, we have
\begin{align}
    -\int\frac{dzp^z}{2\pi}e^{ixzp^z}\ln (z^2 \mu^2 e^{\gamma_E})=&-\int\frac{dzp^z}{2\pi}e^{ixzp^z}\ln (\frac{z^2 \mu^2 e^{\gamma_E}}{K^2})-\int\frac{dzp^z}{2\pi}e^{ixzp^z}\ln K^2\nonumber\\
    =& \tilde{f}^C(x) - \ln K^2\left[\frac{1}{2}\left(\frac{1}{x^2}\delta^+\left(\frac{1}{x}\right)+\frac{1}{(-x)^2}\delta^+\left(-\frac{1}{x}\right)\right)\right] ,\label{A8}
\end{align}
with $K$ a constant.
Hence the integral has a $\delta$ function ambiguity at infinite $|x|$. Although we argue in the next subsection that these $\delta$ functions do not contribute in the matching formula. It is something worth  noticing. 
\subsection{$\delta$ function at infinite $\xi$}
\label{A3}
In this section, we show the $\delta$ functions at infinite $|\xi|$ in the kernel actually do not contribute in the matching. The first case is the ambiguity in the kernel shown in Eq.(\ref{A8}), which has the structure
\begin{align}
\delta C(\xi)=\frac{1}{2}\left[\frac{1}{\xi^2}\delta^+\left(\frac{1}{\xi}\right)+\frac{1}{(-\xi)^2}\delta^+\left(-\frac{1}{\xi}\right)\right] .\label{A9}
\end{align}
Its contribution to the matching is
\begin{align}
   \int\frac{dy}{|y|} \delta C(\frac{x}{y}) q(y)= \lim_{\beta\rightarrow0^+}\int\frac{dy}{|y|}\frac{y^2}{2x^2}\left[\delta\left(\frac{y}{x}-\beta\right)+\delta\left(-\frac{y}{x}-\beta\right)\right]q(y)=\lim_{\beta\rightarrow0^+}\frac{\beta}{2}\left[q(\beta x)+q(-\beta x)\right]=0 ,
   \label{A10}
\end{align}
where in the last equality we have used the fact that the net quark number $\int d x q(x)$ is finite, therefore, $x[q(x)+q(-x)]\propto x^a $ with $a > 0$ as $x \to 0$. Note that Ref.\cite{Izubuchi:2018srq} also asserted that these $\delta$ functions do not contribute to matching because $\lim_{\beta\rightarrow0^+}\beta q(\beta x)=0$. However, this is not satisfied when the sea quarks have infinite number of quarks and antiquarks but with the net quark number to be zero, which is the case from global fits. But Eq.(\Ref{A10}) only requires the net quark number in the hadron is finite.

The second case is the the plus function at infinite $|\xi|$
\begin{align}
\label{A11}
\delta C(\xi)=\p{\frac{1}{\xi}}{1}{\infty}{\infty}+\p{-\frac{1}{\xi}}{-\infty}{-1}{-\infty} ,
\end{align}
whose contribution to the matching is 
\begin{align}
    &\int\frac{dy}{|y|}\delta C\left(\frac{x}{y}\right) q(y)\nonumber\\
    =&\int^0_x\frac{dy}{|y|}\frac{y}{x}q(y)-\int^{-x}_0\frac{dy}{|y|}\frac{y}{x}q(y)+\lim_{\beta\rightarrow0^+}\ln\beta\int\frac{dy}{|y|}\frac{y^2}{x^2}\left[\delta\left(\frac{y}{x}-\beta\right)+\delta\left(-\frac{y}{x}-\beta\right)\right]q(y)\nonumber\\
    =&\int^0_x\frac{dy}{|y|}\frac{y}{x}q(y)-\int^{-x}_0\frac{dy}{|y|}\frac{y}{x}q(y) ,
\end{align}
where we have used $\lim_{\beta\rightarrow0^+}\frac{\beta\ln\beta}{2}\left[q(\beta x)+q(-\beta x)\right]\propto \beta^a \ln\beta \to 0$ with $a > 0$. 
So for the plus function of Eq.(\ref{A11}), we only need to keep the $1/|\xi|$ part. The delta function part can be dropped without any effect in the matching.  
\subsection{Derivation of Eq.(\ref{28})
}\label{A4X}
To derive from Eq.(\ref{31}) to Eq.(\ref{28}), we have the integral 
\begin{align}
C^{\text{hybrid-ratio}}-C^{\text{ratio}}=&\int^{+\infty}_{-\infty}\frac{P^zdz}{2\pi}e^{i(\xi-1) P^zz}\frac{\alpha_sC_F}{2\pi}\frac{3}{2}\ln{\frac{z^2}{y^2z_s^2}}\theta(|z/y|-z_s),
\label{hrci}
\end{align}
which is not well defined for the integration near $z \to \infty$. We use the regulator introduced in Eq.(\ref{A7}) and rewrite $\theta(|z/y|-z_s)=1-\theta(z_s-|z/y|)$ to separate the original integral into two. The first integral is similar to Eq.(\ref{A7}). The second integral yields, 
\begin{align}
&\int^{+\infty}_{-\infty}\frac{P^zdz}{2\pi}e^{ix P^zz}\ln{\frac{z^2}{y^2z_s^2}}(\theta(z_s-|z/y|))\nonumber\\
=&\left[\frac{d}{d\eta}\int^{+ |y|z_s}_{-|y|z_s}\frac{P^zdz}{2\pi}e^{ix P^zz}(\frac{z^2}{y^2z_s^2})^\eta\right]\Bigg\arrowvert_{\eta=0}\nonumber
\\
=&-\frac{1}{\pi}\bigg\{\frac{2\text{Si}( |y|P^zz_sx)}{x}-\lim_{\beta\rightarrow 0^+}\frac{1}{2}\ln\frac{ z_s^2y^2P^2_z}{e^{-2\gamma_E}}\left(\frac{\sin\left(|y|P^zz_s\left(x-\frac{1}{\beta}\right)\right)}{x-\frac{1}{\beta}}+\frac{\sin\left(|y|P^zz_s\left(x+\frac{1}{\beta}\right)\right)}{x+\frac{1}{\beta}}\right)\nonumber
\\
&+\left(\frac{\sin\left(|y|P^zz_s\left(x-\frac{1}{\beta}\right)\right)}{x-\frac{1}{\beta}}+\frac{\sin\left(|y|P^zz_s\left(x+\frac{1}{\beta}\right)\right)}{x+\frac{1}{\beta}}\right)\ln\beta\bigg\}\nonumber\\
=&-\frac{2\text{Si}(|y|P^zz_sx)}{\pi x}+\ln\frac{ z_s^2y^2P^2_z}{e^{-2\gamma_E}}\frac{1}{2x^2}\left(\delta^+\left(\frac{1}{x}\right)+\delta^+\left(-\frac{1}{x}\right)\right)-
\begin{cases}
\left[\frac{1}{x}\right]^{[1,+\infty]}_{+(+\infty)}-\frac{1}{x} & x>1\\
\left[-\frac{1}{x}\right]^{[-\infty,-1]}_{+(-\infty)}+\frac{1}{x} & x<-1 \ . \\
\end{cases}
\end{align}

The two integrals yield the combined result: 
\begin{align}
C^{\text{hybrid-ratio}}-C^{\text{ratio}}=&\frac{\alpha_sC_F}{2\pi}\frac{3}{2}\Bigg\{\ln\frac{e^{-2\gamma_E}}{ z_s^2y^2P^2_z}\delta(1-\xi)+\frac{2\text{Si}((1-\xi)|y|P^zz_s)}{\pi(1-\xi)}-
 \begin{cases}
  \frac{1}{\xi}-\left(\frac{1}{\xi}\right)^{[1,+\infty]}_{+(1)}-\left(\frac{1}{1-\xi}\right)^{[1,+\infty]}_{+(1)}  & \xi > 1\\
  \left(\frac{1}{1-\xi}\right)^{[0,1]}_{+(1)} & 0 < \xi < 1\\
  \frac{1}{1-\xi} & \xi < 0
 \end{cases}
\Bigg\}\nonumber\\
=&\frac{\alpha_sC_F}{2\pi}\frac{3}{2}\left(-\frac{1}{|1-\xi|}+\frac{2\text{Si}((1-\xi)|y|P^zz_s)}{\pi(1-\xi)}\right)^{[-\infty,\infty]}_{+(1)} ,
\end{align}
where we have used the following identity to form the plus function
\begin{align}
\int^\infty_{-\infty} d\xi\left[\frac{2\text{Si}((1-\xi)|y|P^zz_s)}{\pi(1-\xi)}-
 \begin{cases}
  \frac{1}{\xi}& \xi > 1\\
  \frac{1}{1-\xi} & \xi < 0
 \end{cases}\right]=&\ln\frac{ z_s^2y^2P^2_z}{e^{-2\gamma_E}} .
\end{align}
\end{widetext}
\bibliography{ref}
\end{document}